\documentclass{article}
\usepackage{amssymb}
\usepackage{amsmath}

\renewcommand{\theequation}{\thesection.\arabic{equation}}
\begin{document}

\begin{flushright}
\begin{minipage}[t]{55mm}
Preprint of P.N. Lebedev Physical Institute, No. 39, 1975
\end{minipage}
\end{flushright}

\vspace{0.5cm}

\begin{center}

{\LARGE Gauge Invariance in Field Theory and Statistical Physics \\
in Operator Formalism}
\\
\vspace{0.5cm}
{\large
I.V. Tyutin\footnote{{\large E-mail: tyutin@lpi.ru}}
\\
I.E. Tamm Department of Theoretical Physics,\\[0pt]
P.N. Lebedev Institute of Physics,\\[0pt]
119991, Leninsky Prospect 53, Moscow, Russia}
\end{center}

\begin{abstract}
We obtain the Ward identities and the gauge-dependence of Green's
functions in non-Abelian gauge theories by using only the
canonical commutation relations and the equations of motion for
the Heisenberg operators. The consideration is applicable to
theories both with and without spontaneous symmetry breaking. We
present a definition of a generalized statistical average which
ensures that the Fourier images of temperature Green's functions
of the Fermionic fields have only even-valued frequencies. This
makes it possible to set up a procedure of gauge-invariant
statistical averaging in terms of the Hamiltonian and the field
operators.
\end{abstract}

\section{Introduction}

The study of the effects of spontaneously-broken gauge theories in
statistical physics \cite{1} has raised the problem of finding a proof of
the gauge-invariance of physical results in gauge statistical physics.
Formally, such a proof can be carried out by analogy with quantum field
theory, by using a representation of statistical averaging with the help of
functional integration, which has been presented in \cite{2}. However, there
exist some specific calculations of physical effects \cite{3} that have an
appearance of being gauge-dependent. This circumstance may cast a shadow on
the applicability of the functional approach to statistical physics. In
particular, one may raise the question as to the validity of a non-local
change of variables in the functional integral for the partition function,
which has to be made in the course of the usual proof of gauge-invariance,
as well as in the process of deriving the Ward identities.

It appears useful, therefore, to deduce the Ward identities and the
gauge-invariance of physical results in the framework of the operator
formalism of quantum field theory, i.e., by using only the Heisenberg
equations of motion and canonical commutation relations. This has been done
in the present article for both field theory (Sections 3, 4) and statistical
physics (Section 5).

It is essential that the construction of a theory requires to sum up a
gauge-invariant Lagrangian, $L_{0}$, not only with a gauge-fixing term, but
also with an additional Lagrangian, $L_{C}$, responsible for an interaction
of fictitious particles with the gauge field. For those gauges that are
usually applied in quantum electrodynamics, the fictitious particle is free,
so that the theory can be set up without the term $L_{C}$. In the case of
non-Abelian gauge theories, it is well-known that there arise some
additional diagrams (with respect to the Feynman diagrams) that effectively
describe an interaction of the gauge field with the fictitious particles. We
choose the Lagrangian $L_{C}$ in such a way that, on the one hand, it
implies the necessary additional diagrams, and, on the other hand, it admits
a canonical quantization.

The deduction of the Ward identities and the proof of
gauge-invariance that are based on the equations of motion for the
Heisenberg fields can also be useful for other purposes, such as a
description of gauge theories in the framework of Zimmermann's normal
product \cite{4}.

The fictitious particles described by the Lagrangian $L_{C}$ are scalars,
but, at the same time, they are Fermions. Thus, the usual definition of
statistical average leads to such a Fourier-image of the
temperature Green function of these particles that contains only odd-valued
frequencies. However, as will be shown in Section 5, the gauge-invariance of
the partition function demands that the Green function of fictitious
particles should contain only even-valued frequencies (whereas the Green
function of Bose particles should contain odd-valued frequencies). This
allows one to pose statistics in terms of an operator formalism by using the
Hamiltonian and the field operators. Physical quantities, and, in
particular, the partition function, prove to be gauge-invariant. In the case
of gauges that eliminate the fictitious particles, the generalized
definition of the partition function becomes identical with the usual
definition.

The present consideration has been made for a sufficiently large class of
gauge conditions and is applicable to theories both with and without
spontaneous symmetry breaking.

\section{General formulas}

In this section, we present some general formulas that we use in the
following sections.

We examine a Lagrangian of a renormalizable theory of a gauge-invariant
interaction of spinless and Fermionic fields with a gauge field of a general
kind:%
\begin{eqnarray}
&&L=L_{0}+\frac{1}{2}t^{a}\alpha _{ab}t^{b}+L_{C}\,,  \label{2.1} \\
&&L_{0}=-\frac{1}{4}G_{\mu \nu }^{a}G^{a,\mu \nu }+\frac{1}{2}\left[ \left(
\partial _{\mu }-ig\Gamma ^{a}A_{\mu }^{a}\right) \varphi \right] ^{2}+
\notag \\
&&+\,\,\overline{\psi }\gamma ^{\mu }\left( i\partial _{\mu }+g\tau ^{a}A_{\mu
}^{a}\right) \psi -V\left( \psi ,\varphi \right) \,,  \label{2.1'}
\end{eqnarray}%
where $\varphi _{i}$ is a multiplet of Hermitian spinless fields; $\psi _{i}$
is a multiplet of Fermionic fields; $V\left( \psi ,\varphi \right) $ is an
arbitrary Lagrangian of interaction of the fields $\psi $, $\varphi $ that
obeys the requirement of gauge-invariance; $G_{\mu \nu }^{a}$ is the
strength tensor of a Yang--Mills field $A_{\mu }^{a}$:%
\begin{equation}
G_{\mu \nu }^{a}=\partial _{\mu }A_{\nu }^{a}-\partial _{\nu }A_{\mu
}^{a}+gf^{abc}A_{\mu }^{b}A_{\nu }^{c}\,,  \label{2.2}
\end{equation}%
while $f^{abc}$ are structure constants of an invariance group $G$; the
matrices $\Gamma ^{a}$ and $\tau ^{a}$ are the generators of the
transformation group for the fields $\varphi _{i}$ and $\psi _{i}$, with the
commutation relations%
\begin{equation}
\left[ \Gamma ^{a},\Gamma ^{b}\right] =if^{abc}\Gamma ^{c}\,,  \label{2.3}
\end{equation}%
and $\tau ^{a}$ obey similar relations. The matrices $\Gamma ^{a}$ are
antisymmetric and have purely imaginary values.

The last two terms in (\ref{2.1}) describe subsidiary conditions
that are necessary for the possibility of a canonical description
of the theory (as well as for the existence of a perturbation
theory). Let us choose the functions $t^{a}$ in the form%
\begin{equation}
t^{a}=\varkappa ^{\mu \nu }\partial _{\mu }A_{\nu }^{a}+\varkappa
_{i}^{a}\varphi_{i}\,,  \label{2.4}
\end{equation}%
where $\varkappa ^{\mu \nu }=\varkappa ^{\nu \mu }$ is a matrix that has
numerical elements; the matrices $\varkappa _{i}^{a}$ will be described
below.

The Lagrangian $L_{C}$ describes an interaction of fictitious particles
(Fermionic scalars), $C^{a}$, $C^{+a}$, with the fields $A_{\mu }^{a}$, $%
\varphi _{i}$ and implies the well-known additional diagrams in quantum
theory \cite{5}, namely,%
\begin{eqnarray}
&&\,L_{C}=-\partial _{\mu }C^{+a}\varkappa ^{\mu \nu }\nabla _{\nu
}^{ab}C^{b}+igC^{+a}\varkappa _{i}^{a}\Gamma _{ij}^{b}\varphi _{j}C^{b}\,,
\label{2.5} \\
&&\,\nabla _{\nu }^{ab}=\partial _{\nu }\delta ^{ab}+gf^{acb}A_{\nu }^{c}\,.
\notag
\end{eqnarray}

The matrix $\varkappa _{i}^{a}$ is chosen as follows: let the vacuum mean
value of the field $\varphi _{i}$ be non-vanishing,%
\begin{equation}
\left\langle 0|\varphi _{i}|0\right\rangle =\xi _{i}\,.  \label{2.6}
\end{equation}%
The set of generators $\Gamma _{ij}^{a}$ can always be split in two groups, $%
\Gamma ^{a}=\left( \Gamma ^{m},\Gamma ^{l}\right) $, such that%
\begin{equation}
\Gamma ^{l}\xi =0,\;\Gamma ^{m}\xi =i\xi ^{m}\not=0\,,  \label{2.7}
\end{equation}%
where the vectors $\xi ^{m}$ are orthogonal. Then $\varkappa _{i}^{a}$ must
obey the conditions%
\begin{equation}
\varkappa _{i}^{a}\xi _{i}=0\,.  \label{2.8}
\end{equation}%
Besides, either (for a fixed $a$)%
\begin{equation}
\varkappa _{i}^{a}\equiv 0\,,  \label{2.9}
\end{equation}%
or%
\begin{equation}
\varkappa _{i}^{a}\xi _{i}^{m}\not=0\,,  \label{2.9'}
\end{equation}%
with a certain $m$. In other respects, the matrix $\varkappa _{i}^{a}$ is
arbitrary. We do not discuss any restrictions that may be imposed on $%
\varkappa _{i}^{a}$ by conditions (\ref{2.8})--(\ref{2.9'}). Notice that
conditions (\ref{2.8}) and (\ref{2.9'}) are necessary only for the
possibility of passing to unitary gauges of the kind $\varkappa
_{i}^{a}\varphi _{i}=0$. By themselves, they are unnecessary to provide the
possibility of a canonical description.

In principle, one can choose a subsidiary condition of a yet more general
kind, being compatible with renormalizability:%
\begin{equation*}
t\sim \partial A+\partial \varphi +A+\varphi +A^{2}+A\varphi +\varphi ^{2}\,.
\end{equation*}%
This, however, will only imply some obvious modifications of the reasonings
to be presented below.

Since we examine a theory which admits a possibility of spontaneous symmetry
breaking, let us now introduce some spinless fields, whose vacuum mean value
is zero:%
\begin{equation}
\varphi _{i}=\xi _{i}+\sigma _{i}\,,\;\left\langle 0|\sigma
_{i}|0\right\rangle =0\,.  \label{2.10}
\end{equation}%
The Lagrangian $L_{0}$ is invariant with respect to gauge transformations
whose infinitesimal form is given by%
\begin{eqnarray}
&&A_{\mu }^{a}\rightarrow A_{\mu }^{\prime a}=A_{\mu }^{a}+\nabla _{\mu
}^{ab}\Lambda ^{b}\,,\;\nabla _{\mu }^{ab}=\partial _{\mu }\delta
^{ab}+gf^{acb}A_{\mu }^{c}\,,  \label{2.11} \\
&&\psi \rightarrow \psi ^{\prime }=\psi +ig\Lambda ^{a}\tau ^{a}\psi
\,,\;\varphi \rightarrow \varphi ^{\prime }=\varphi +ig\Lambda ^{a}\Gamma
^{a}\varphi \,,  \label{2.11'}
\end{eqnarray}%
where $\Lambda ^{a}$ are infinitesimal parameters of the gauge
transformations; they depend on the coordinates. For the field $\sigma $,
transformation (\ref{2.11'}) reads as follows:%
\begin{equation}
\sigma \rightarrow \sigma ^{\prime }=\sigma +ig\Lambda ^{a}\Gamma ^{a}\xi
+ig\Lambda ^{a}\Gamma ^{a}\sigma \,,  \label{2.11''}
\end{equation}%
i.e., $\sigma $ transforms in a non-homogenous manner. The invariance of $%
L_{0}$ with respect to (\ref{2.11})--(\ref{2.11''}) implies the identities%
\begin{equation}
-\nabla _{\mu }^{ab}\frac{\delta L_{0}}{\delta A_{\mu }^{b}}+ig\frac{\delta
L_{0}}{\delta \sigma _{i}}\Gamma _{ij}^{a}\varphi _{j}-ig\frac{\delta L_{0}}{%
\delta \psi _{i}}\tau _{ij}^{a}\psi _{j}-ig\overline{\psi }_{i}\tau _{ij}^{a}%
\frac{\delta L_{0}}{\delta \overline{\psi }_{j}}\equiv 0\,.  \label{2.12}
\end{equation}

Let us now proceed to the canonical quantization of the theory. The momenta
of the fields $A_{\mu }^{a}$, $\varphi $ and $\psi $ are defined as usual:%
\begin{eqnarray}
&&\pi _{A}:\;\pi ^{a,0}=\alpha _{ab}\varkappa ^{00}t^{b}\,,\;\pi
^{a,k}=-G^{a,0k}+\alpha _{ab}\varkappa ^{0k}t^{b}\,,  \label{2.13} \\
&&\Pi _{\sigma }:\;\Pi _{i}=\dot{\sigma}_{i}-ig\Gamma
_{ik}^{a}A_{0}^{a}\varphi _{k}\,,  \label{2.13'} \\
&&\Pi _{\psi }:\;\overline{\Pi }_{i}=i\psi _{i}^{+}\,,  \label{2.13''}
\end{eqnarray}

Using these relations, one can obtain the derivatives of the fields with
respect to time:%
\begin{eqnarray}
&&t^{a}=\frac{1}{\varkappa ^{00}}\alpha ^{ab}\pi ^{b,0}\,,\;\alpha
^{ab}\alpha _{bc}=\delta _{c}^{a}\,,  \label{2.14} \\
&&G^{a,0k}=-\pi ^{a,k}+\frac{1}{\varkappa ^{00}}\varkappa ^{0k}\pi ^{a,0}\,,
\notag \\
&&\dot{A}^{a,k}=\frac{\varkappa ^{0k}}{\varkappa ^{00}}\pi ^{a,0}-\pi
^{a,k}+\nabla ^{ab,k}A^{b,0}\,,  \label{2.14'} \\
&&\,\dot{A}_{0}^{a}=\frac{1}{\varkappa ^{00}}\left( \frac{1}{\varkappa ^{00}}%
\alpha ^{ab}\pi ^{b,0}-\varkappa ^{\mu i}\partial _{i}A_{\mu }^{a}-\varkappa
_{i}^{a}\sigma _{i}+\varkappa _{0k}\pi ^{a,k}-\right.  \label{2.14''} \\
&&\,\left. -\frac{\varkappa _{0k}\varkappa ^{0k}}{\varkappa ^{00}}\pi
^{a,0}-\varkappa ^{0k}\nabla _{k}^{ab}A_{0}^{b}\right) \,,  \notag \\
&&\,\dot{\sigma}=\Pi _{i}+ig\Gamma _{ij}^{a}A_{0}^{a}\varphi _{j}\,.
\label{2.14'''}
\end{eqnarray}

The canonical variables obey the usual equal-time commutation relations%
\begin{equation}
\left[ A_{\mu }^{a},\pi ^{b,\nu }\right] =i\delta ^{ab}\delta _{\nu }^{\mu
}\,,\;\left[ \sigma _{i},\Pi _{j}\right] =i\delta _{ij}\,,\;\left[ \psi _{i},%
\overline{\Pi }_{j}\right] =i\delta _{ij}\,.  \label{2.15}
\end{equation}

To find the canonical momenta conjugate to the fictitious fields, as well as
the corresponding anticommutation relations, one can use Schwinger's action
principle \cite{6}; see Appendix A, where it has also been demonstrated that
these expressions are, in fact, the only possible ones. As a result, we find%
\begin{eqnarray}
&&\Pi _{C}^{a}=-\varkappa ^{0\mu }\partial _{\mu }C^{+a}\,,\;\Pi
_{C^{+}}^{a}=\varkappa ^{0\mu }\nabla _{\mu }^{ab}C^{b}\,,  \label{2.16} \\
&&\dot{C}^{+a}=-\frac{1}{\varkappa ^{_{00}}}\left( \Pi _{C}^{a}+\varkappa
^{0k}\partial _{k}C^{+a}\right) \,,  \notag \\
&&\dot{C}^{a}=\frac{1}{\varkappa ^{_{00}}}\left( \Pi _{C^{+}}^{a}-g\varkappa
^{00}f^{abd}A_{0}^{b}C^{d}-\varkappa ^{0k}\nabla _{k}^{ab}C^{b}\right) \,,
\label{2.17} \\
&&\,\{C^{a},\Pi _{C}^{b}\}=\{C^{+a},\Pi _{C^{+}}^{b}\}=i\delta ^{ab}\,.
\label{2.18}
\end{eqnarray}%
The other anticommutators are equal to zero. The corresponding Hamiltonian
reads%
\begin{eqnarray}
&&\,H=\pi ^{a,\mu }\dot{A}_{\mu }^{a}+\Pi _{i}\dot{\sigma}_{i}+\overline{\Pi
}_{i}\dot{\psi}_{i}+\Pi _{C}^{a}\dot{C}^{a}+\Pi _{C^{+}}^{a}\dot{C}^{+a}-L=
\notag \\
&&\,=\frac{1}{2\left( \varkappa ^{00}\right) ^{2}}\pi ^{a,0}\left( \alpha
^{ab}-\delta _{ab}\varkappa _{0k}\varkappa ^{0k}\right) \pi ^{b,0}-\frac{1}{2%
}\pi ^{a,k}\pi _{a,k}+\pi ^{a,k}\nabla _{k}^{ab}A_{0}^{b}+  \notag \\
&&\,+\,\,\frac{1}{\varkappa ^{00}}\pi ^{a,0}\left( \varkappa ^{0k}\pi
_{k}^{a}-\varkappa ^{i\nu }\partial _{i}A_{\nu }^{a}-\varkappa ^{0k}\nabla
_{k}^{ab}A_{0}^{b}-\varkappa _{i}^{a}\sigma _{i}\right) +\frac{1}{2}\Pi
_{i}^{2}+  \notag \\
&&+\,\,ig\Pi _{i}\Gamma _{ik}^{a}A_{0}^{a}\varphi _{k}+\frac{1}{4}%
G_{ik}^{a}G^{a,ik}-\frac{1}{2}\left( \partial _{k}-ig\Gamma
^{a}A_{k}^{a}\right) \varphi \cdot \left( \partial ^{k}-ig\Gamma
^{b}A^{b,k}\right) \varphi -  \notag \\
&&\,-\,\,\overline{\psi }\gamma ^{k}\left( i\partial _{k}+g\tau
^{a}A_{k}^{a}\right) \psi -g\overline{\psi }\gamma ^{0}\tau
^{a}A_{0}^{a}\psi +\frac{1}{\varkappa ^{00}}\Pi _{C}^{a}\Pi _{C^{+}}^{a}-
\notag \\
&&-\,\,\frac{1}{\varkappa ^{00}}\Pi _{C}^{a}\varkappa ^{0k}\nabla
_{k}^{ab}C^{b}-\frac{1}{\varkappa ^{00}}\Pi _{C^{+}}^{a}\varkappa
^{0k}\partial _{k}C^{+a}-g\Pi _{C}^{a}f^{abd}A_{0}^{b}C^{d}+  \notag \\
&&\,+\,\,\partial _{i}C^{+a}\left( \varkappa ^{ij}-\frac{\varkappa
^{0i}\varkappa ^{0j}}{\varkappa ^{00}}\right) \nabla
_{j}^{ab}C^{b}-igC^{+a}\varkappa _{i}^{a}\Gamma _{ij}^{b}\varphi _{j}C^{b}\,.
\label{2.19}
\end{eqnarray}%
One can easily see that the canonical equations that follow from Hamiltonian
(\ref{2.19}) are identical with the Lagrangian equations. For the fictitious
particles, this has been verified in Appendix A.

We need an expression for $\dot{\pi}^{a,0}$ in terms of the canonical
coordinates and momenta. It can be found with the help of the following
relation, that holds true for any operator $Q$:%
\begin{equation}
\dot{Q}=i\left[ H,Q\right] \,.  \label{2.20}
\end{equation}%
We have%
\begin{eqnarray}
&&\dot{\pi}^{a,0}=-\frac{1}{\varkappa ^{00}}\varkappa ^{0i}\partial _{i}\pi
^{a,0}-\frac{1}{\varkappa ^{00}}\varkappa ^{0i}\nabla _{i}^{ab}\pi
^{b,0}+\nabla _{k}^{ab}\pi ^{b,k}-  \notag \\
&&-\,\,ig\Pi _{i}\Gamma _{ij}^{a}\varphi _{j}+g\overline{\psi }\gamma
^{0}\tau ^{0}\psi +g\Pi _{C}^{b}f^{bad}C^{d}\,.  \label{2.21}
\end{eqnarray}%
From (\ref{2.12}), which holds identically, it follows that on the equations
of motion relation (\ref{2.12}) is valid also for the quantity $L-L_{0}$,
that is,%
\begin{eqnarray}
&&\,\left( \nabla _{\mu }^{ab}\varkappa ^{\mu \nu }\partial _{\nu
}+ig\varkappa _{i}^{b}\Gamma _{ij}^{a}\varphi _{j}\right) \alpha
_{bc}t^{c}\equiv t^{b}\alpha _{bc}T^{ca}=  \notag \\
&&\,=-g\varkappa ^{\mu \nu }\partial _{\mu }\left( \partial _{\nu
}C^{+b}f^{bad}C^{d}\right) -g^{2}\hat{A}_{\mu }^{ab}\varkappa ^{\mu \nu
}\partial _{\nu }C^{+d}f^{dbf}C^{f}+  \notag \\
&&\,+\,\,g^{2}C^{+b}\varkappa _{i}^{b}\Gamma _{ij}^{d}\Gamma
_{jk}^{a}\varphi _{k}C^{d}\,,\;\hat{A}_{\mu }^{ab}=f^{acb}A_{\mu }^{c}\,,
\label{2.22}
\end{eqnarray}%
where $T^{ab}$ stands for the operator%
\begin{equation}
T^{ab}=\partial _{\mu }\varkappa ^{\mu \nu }\nabla _{\nu }^{ab}+ig\varkappa
_{i}^{a}\Gamma _{ij}^{b}\varphi _{j}\,.  \label{2.23}
\end{equation}%
The equations of motion for the fictitious fields read as follows:%
\begin{equation}
T^{ab}C^{b}=C^{+b}T^{ba}\,.  \label{2.24}
\end{equation}

\section{Ward identities}

\setcounter{equation}{0}

In this section, we deduce the Ward identities by using only the equations
of motion and canonical commutation relations.

Consider the vacuum mean value%
\begin{equation}
Z_{C^{d}C^{+a}}\equiv \left\langle 0\left| TC^{d}\left( y\right)
C^{+a}\left( x\right) \exp \left( iQ\right) \right| 0\right\rangle \,,
\label{3.1}
\end{equation}%
where%
\begin{equation}
Q=\int du\,Q\left( u\right) =\int du\,\left( J^{b,\mu }A_{\mu
}^{b}+J_{i}\varphi _{i}+\overline{\eta }\psi +\overline{\psi }\eta \right)
\,.  \label{3.2}
\end{equation}

An arbitrary operator $P\left( z\right) $ obeys the relation%
\begin{eqnarray}
&&\,\partial _{0}^{\left( z\right) }\left\langle P\left( z\right)
\right\rangle =\left\langle \dot{P}\left( z\right) \right\rangle +i\int
du\,\delta \left( z_{0}-u_{0}\right) \left\langle \left[ P\left( z\right)
,Q\left( u\right) \right] \right\rangle +  \notag \\
&&\,+\left\langle 0\left| T\left\{ \exp \left( iQ\right) \delta \left(
z_{0}-y_{0}\right) \left[ P\left( z\right) ,C^{d}\left( y\right) \right]
C^{+a}\left( x\right) \right. \right. \right. +  \notag \\
&&\,+\left. \left. \left. C^{d}\left( y\right) \delta \left(
z_{0}-x_{0}\right) \left[ P\left( z\right) ,C^{+a}\left( x\right) \right]
\right\} \right| 0\right\rangle \,,  \label{3.3}
\end{eqnarray}%
where the following notation has been used:%
\begin{equation}
\left\langle P\left( z\right) \right\rangle \equiv \left\langle 0|TP\left(
z\right) C^{d}\left( y\right) C^{+a}\left( x\right) \exp \left( iQ\right)
|0\right\rangle \,.  \label{3.4}
\end{equation}%
Because of the fact that all the fields (anti)commute at equal times, we have%
\begin{equation}
t^{f}\left( z\right) Z_{CC^{+}}=\left\langle t^{f}\left( z\right)
\right\rangle \,.  \label{3.5}
\end{equation}

Expressions of the form $P\left( A,\varphi ,\psi ,\overline{\psi }\right)
Z_{CC^{+}}$ imply that the function $P$ is subject to the replacement%
\begin{equation}
A\rightarrow \frac{1}{i}\frac{\delta }{\delta J}\,,\;\mathrm{etc\,}.
\label{3.6}
\end{equation}%
In particular,%
\begin{equation}
t^{f}\left( z\right) Z_{CC^{+}}=\varkappa ^{\mu \nu }\partial _{\mu
}^{\left( z\right) }\left\langle A^{f}\left( z\right) \right\rangle
+\varkappa _{i}^{f}\left\langle \varphi _{i}\left( z\right) \right\rangle \,.
\label{3.7}
\end{equation}%
Further, relations (\ref{2.13})--(\ref{2.18}), (\ref{2.21}), (\ref{3.3})
lead to%
\begin{eqnarray}
&&\overset{\mathbf{.}}{t}^{f}\left( z\right) Z_{CC^{+}}=\left\langle \dot{t}%
^{f}\left( z\right) \right\rangle +\frac{1}{\varkappa ^{00}}\alpha
^{fb}J^{b,0}\left( z\right) Z_{CC^{+}}\,\,,  \label{3.8} \\
&&\overset{\mathbf{..}}{t}^{f}\left( z\right) Z_{CC^{+}}=\left\langle
\overset{\mathbf{..}}{t}^{f}\left( z\right) \right\rangle +\frac{1}{%
\varkappa ^{00}}\alpha ^{fb}\dot{J}^{b,0}\left( z\right) Z_{CC^{+}}-\frac{1}{%
\varkappa ^{00}}\alpha ^{fb}\left\langle Q_{R}^{b}\left( z\right)
\right\rangle -  \notag \\
&&-\,\,\frac{1}{\varkappa ^{00}}\alpha ^{fb}\left\langle \left[ \frac{1}{%
\varkappa ^{00}}\varkappa ^{0i}\partial _{i}Z^{b,0}\left( z\right) +\frac{1}{%
\varkappa ^{00}}\varkappa ^{0i}\nabla _{i}^{bc}J^{c,0}\left( z\right)
+\nabla _{0}^{bc}J^{c,0}\left( z\right) \right] \right\rangle +  \notag \\
&&+\,\,\delta \left( z-y\right) \frac{1}{\varkappa ^{00}}\alpha
^{fb}f^{bdd^{\prime }}Z_{C^{d^{\prime }}C^{+a}}\,\,.  \label{3.9}
\end{eqnarray}%
In (\ref{3.9}), the expression $Q_{R}$ denotes an infinitesimal variation of
the term with the sources,%
\begin{equation}
Q_{R}^{b}\left( z\right) =-\nabla _{\mu }^{bc}J^{c,\mu }\left( z\right)
+igJ_{i}\left( z\right) \Gamma _{ij}^{b}\varphi _{j}\left( z\right) +ig%
\overline{\eta }\left( z\right) \tau ^{b}\psi \left( z\right) -ig\overline{%
\psi }\left( z\right) \tau ^{b}\eta \left( z\right) \,.  \label{3.10}
\end{equation}%
Relations (\ref{3.8}) and (\ref{3.9}) allow one to obtain the following:%
\begin{eqnarray}
&&t^{f}\left( z\right) \alpha _{fc}T^{cb}Z_{CC^{+}}=\left\langle t^{f}\left(
z\right) \alpha _{fc}T^{cb}\right\rangle -  \notag \\
&&-\left\langle Q_{R}^{b}\left( z\right) \right\rangle +i\delta \left(
z-y\right) f^{bdd^{\prime }}Z_{C^{d^{\prime }}C^{+d}}\,.  \label{3.11}
\end{eqnarray}%
We now use relation (\ref{2.22}) in the first summand of the r.h.s. of (\ref%
{3.11}), which is a valid operation, since all the derivatives are already
under the symbol of $T$-product; we assume $b=d$, $y=z$ and take an integral
over $y$, as well as a sum over $d$, namely,%
\begin{eqnarray}
&&\hspace{-1.5cm}\int dy\,t^{f}\left( y\right) \alpha
_{fc}T^{cd}Z_{CC^{+}}=-\int dy\left\langle Q_{R}^{d}\left( y\right)
\right\rangle +  \notag \\
&&\hspace{-1.5cm}+\int dy\left\langle \left[ -g\varkappa ^{\mu \nu }\partial
_{\mu }\left( \partial _{\nu }C^{+b}\left( y\right) f^{bdf}C^{f}\left(
y\right) \right) -g^{2}\hat{A}_{\mu }^{db}\left( y\right) \varkappa ^{\mu
\nu }\partial _{\nu }C^{+f}\left( y\right) f^{fbn}C^{n}\left( y\right)
+\right. \right.  \notag \\
&&\hspace{-1.5cm}+\left. \left. g^{2}C^{+b}\left( y\right) \varkappa
_{i}^{b}\Gamma _{ij}^{f}\Gamma _{jk}^{d}\varphi _{k}\left( y\right)
C^{f}\left( y\right) \right] \right\rangle \,.  \label{3.12}
\end{eqnarray}%
Using the equations of motion for $C$ and $C^{+}$ (\ref{2.24}), as well as
the anticommutativity of the operators $C$, one can prove (see Appendix B)
that the second summand in the r.h.s. of (\ref{3.12}) is equal to zero. Then
(\ref{3.12}) takes the form%
\begin{equation}
\int dy\,t^{b}\left( y\right) \alpha _{bc}T^{cd}Z_{CC^{+}}=-\left\langle
\int dyQ_{R}^{d}\left( y\right) \right\rangle \,.  \label{3.13}
\end{equation}%
In view of expression (\ref{2.23}), one can easily see that the derivatives
in all the terms $T^{cd}$ commute with the symbol of $T$-product, with the
exception of the term $\partial _{0}\varkappa ^{0\nu }\nabla _{\nu }$. By
virtue of (\ref{2.24}), (\ref{2.16}), (\ref{2.18}), we find%
\begin{equation}
T^{cd}Z_{CC^{+}}=i\delta ^{cd}\delta \left( y-z\right) Z\,,  \label{3.14}
\end{equation}%
where $Z$ stands for the vacuum mean value%
\begin{equation}
Z\equiv \left\langle 0\left\vert T\exp \left( iQ\right) \right\vert
0\right\rangle \,.  \label{3.15}
\end{equation}%
As a result, the Ward identity for the function $Z$ takes the form%
\begin{equation}
\alpha _{ab}t^{b}\left( x\right) =i\left\langle 0\left\vert T\int
dyQ_{R}^{b}\left( y\right) C^{b}\left( y\right) C^{+a}\left( x\right)
\right\vert 0\right\rangle \,.  \label{3.16}
\end{equation}%
In order to present (\ref{3.16}) in the usual form, we utilize a relation that
follows from (\ref{3.14}),%
\begin{equation}
P\left( A,\varphi ,\psi \right) Z_{CC^{+}}=iD^{da}\left( y,x\right) P\left(
A,\varphi ,\psi \right) Z\,,\;T^{ab}D^{bc}=\delta ^{ac}\,,  \label{3.17}
\end{equation}%
for an arbitrary function $P$. This relation allows one to present (\ref%
{3.16}) in the usual form which can be found in the literature:%
\begin{equation}
\left[ \alpha _{ab}t^{b}\left( x\right) +\int dy\,Q_{R}^{b}\left( y\right)
D^{ba}\left( y,x\right) \right] Z=0\,.  \label{3.18}
\end{equation}%
For the first time, an identity of the form (\ref{3.18}) has been obtained
by Fradkin \cite{7} for an Abelian theory, as well as by Slavnov \cite{8}
and Taylor \cite{9} for a non-Abelian gauge theory.

Identity (\ref{3.16}) has a simple meaning. The substitution of a new field $%
t\left( x\right) $ into this Green function is equivalent to the sum (over
the number of fields in the initial Green function) of Green's functions
that do not contain the field $t\left( x\right) $ and are deduced from the
initial Green function with the help of an infinitesimal gauge
transformation (\ref{2.11})--(\ref{2.11''})\ of one of the fields with the
gauge function $\Lambda \sim CC^{+}$. In case one of the fields in the Green
function is a gauge-invariant operator, an insertion of the field $t\left(
x\right) $ into such a function yields zero.

\section{Gauge-dependence of Green's functions}

\setcounter{equation}{0}

In this section, we find a relation between Green's functions in the gauges $%
\alpha _{ab}$, $\varkappa ^{\mu \nu }$, $\varkappa _{i}^{a}$ and%
\begin{equation*}
\overline{\alpha }_{ab}=\alpha _{ab}+\delta \alpha _{ab}\,,\;\overline{%
\varkappa }^{\mu \nu }=\varkappa ^{\mu \nu }+\delta \varkappa ^{\mu \nu
}\,,\;\overline{\varkappa }_{i}^{a}=\varkappa _{i}^{a}+\delta \varkappa
_{i}^{a}\,.
\end{equation*}

As a preliminary step, we make the following remark. In the framework of
canonical formalism, the generating functionals (\ref{3.1}), (\ref{3.15}),
or, equivalently, the Green function, are computed by making use of the
vertices determined by $H_{\mathrm{int}}$ and also with the help of
non-covariant propagators, being the actual vacuum mean values of the $T$%
-products of free fields.

In case the Hamiltonian is quadratic in its momenta, one can pass (due to
Wick) to an effective diagrammatic technique,\footnote{%
One can pass to this diagrammatic technique also in the general case.
However, in case the Hamiltonian is more than quadratic in its momenta, one
cannot find the additional diagrams in a manifest form.} in which the
vertices are determined by\ $-L_{\mathrm{int}}$; the field propagators are
determined by $\Lambda ^{-1}$ ($\Lambda $ being a differential operator that
enters the free Lagrangian $L\sim \frac{1}{2}\varphi \Lambda \varphi $),
and, besides, there may also appear some additional vertices.

The additional vertices are determined by the matrix present in those
summands of the Hamiltonian that are quadratic in momenta \cite{9}. In the
case under consideration, the effective Lagrangian that determines the
additional diagrams equals to%
\begin{equation}
-\frac{i}{2}\delta \left( 0\right) \mathrm{Sp\,}\ln \alpha _{ab}\,,  \label{4.1}
\end{equation}%
i.e., there are no additional diagrams. Thus, the generating functionals (%
\ref{3.1}), (\ref{3.15}) can be presented in the form%
\begin{eqnarray}
&&Z=\exp \left( \frac{1}{2}\delta \left( 0\right) \mathrm{Sp\,}\ln \alpha
_{ab}\right) Z_{W}\,,  \label{4.2} \\
&&Z_{CC^{+}}=\exp \left( \frac{1}{2}\delta \left( 0\right) \mathrm{Sp\,}\ln
\alpha _{ab}\right) Z_{CC^{+}W}\,,  \label{4.2'}
\end{eqnarray}%
where $Z_{W}$ and $Z_{CC^{+}W}$ are now computed with the help of the usual
Feynman diagrams.

It is easy to demonstrate (Appendix C) that the variation of Green's
functions with respect to the parameters entering the Lagrangian $L$ is
given by an insertion of the ``field'' $i\int dx\,\delta L$ into Green's
functions. Due to the fact that the vacuum\ mean value $\xi $ depends on the
gauge parameters, we have%
\begin{equation}
\delta Z=i\int dx\,\overline{\delta }LZ+i\int dx\,\delta \xi _{i}\left(
\frac{\delta L}{\delta \sigma _{i}}+\frac{\delta Q}{\delta \sigma _{i}}%
\right) Z+\frac{1}{2}\delta \left( 0\right) \alpha ^{ab}\delta \alpha
_{ab}Z\,,  \label{4.3}
\end{equation}%
where $\overline{\delta }L$ denotes a variation of the Lagrangian with
respect to those gauge parameters that enter the Lagrangian in a manifest
way:%
\begin{equation}
\int dx\,\overline{\delta }L=i\int dx\,\left[ t^{a}\left( \frac{1}{2}\alpha
^{bc}\delta \alpha _{cd}t^{d}+\delta t^{b}\right) \alpha _{ab}+C^{+a}\delta
T^{ab}C^{b}\right] \,.  \label{4.4}
\end{equation}%
The second term in (\ref{4.3}) equals to zero, because it represents one of
the equations satisfied by $Z$ as a function of the sources. This fact
follows from the Euler equations for the fields, as well as from definition (%
\ref{3.15}) and from the canonical commutation relations.

Let us now use the Ward identity in the form (\ref{3.16}):%
\begin{eqnarray}
&&\hspace{-1cm}\delta Z=-\int dx\,\left( \frac{1}{2}\alpha ^{ac}\delta
_{cb}t^{b}\left( x\right) +\delta t^{a}\left( x\right) \right) \left\langle
0\left| T\exp \left( iQ\right) \int dy\,Q_{R}^{d}\left( y\right) C^{d}\left(
y\right) C^{+a}\left( x\right) \right| 0\right\rangle +  \notag \\
&&\hspace{-1cm}+\,\,i\left\langle 0\left| T\exp \left( iQ\right) \int
dx\,C^{+a}\left( x\right) T^{ab}C^{b}\left( x\right) \right| 0\right\rangle +%
\frac{1}{2}\delta \left( 0\right) \alpha ^{ab}\delta \alpha _{ab}Z\,.
\label{4.5}
\end{eqnarray}%
With allowance for definitions (\ref{2.4}), (\ref{2.23}) and (\ref{3.10}),
the sum of the second term and the part of the first term which contains $%
\delta t$ in the r.h.s. of (\ref{4.5}) yields%
\begin{equation}
-\left\langle 0\left| T\exp \left( iQ\right) \int dxdy\,Q_{R}^{a}\left(
x\right) C^{a}\left( x\right) C^{+b}\left( y\right) \delta t^{b}\left(
y\right) \right| 0\right\rangle \,.  \label{4.6}
\end{equation}%
Let us recall that we compute $Z$ with the help of Feynman diagrams, so that
the time derivatives of $t$, $\delta t$, $\delta T$ and $Q_{R}$ should be
regarded as commuting with the symbol of $T$-product. However, relation (\ref%
{3.17}) remains valid. Due to this fact, we finally have%
\begin{eqnarray}
&&\delta Z=-i\int dxdy\,Q_{R}^{a}\left( x\right) D^{ab}\left( x,y\right)
\overline{\Lambda }^{b}\left( y\right) Z\,,  \label{4.7} \\
&&\Lambda ^{a}\left( x\right) =\frac{1}{2}\alpha ^{ab}\delta \alpha
_{bc}t^{c}\left( x\right) +\delta t^{c}\left( x\right) \,.  \label{4.8}
\end{eqnarray}

Relation (\ref{4.7}) has a simple meaning as well:\ an infinitesimal change
of the gauge parameters in the Green function is equivalent to an
infinitesimal gauge transformation of each of the fields with the gauge
parameter $\Lambda \sim D\overline{\Lambda }$. In particular,
(non-renormalized) Green's functions of gauge-invariant operators are
gauge-independent.

Concluding this section, notice that one could remain in the framework of
the Dyson $T$-product and use the variation $\delta Z$ (Appendix C) in the
form%
\begin{equation*}
\delta Z=-i\left\langle 0\left| T\int dx\,\delta H\exp \left( iQ\right)
\right| 0\right\rangle \,.
\end{equation*}%
This would only lead to some more tedious calculations, that would naturally
leave relation (\ref{4.7}) unaltered.

\section{Gauge invariance of partition function}

\setcounter{equation}{0}

In this section, we examine the problem of defining the partition function
in gauge theories.

Notice that if one defines the partition function as usual,%
\begin{equation}
Z=\mathrm{Sp\,}e^{-\beta H}  \label{5.1}
\end{equation}%
[where $\mathrm{Sp}$ is defined (in a space with indefinite metric) so that
a cyclic permutation of operators under the symbol of $\mathrm{Sp}$ is
admissible], then it is not gauge-invariant, since this case does not comply
with the Ward identities (\ref{3.16}), which provide a basis for the
gauge-invariance of physical quantities.

Indeed, in the simplest case of free electromagnetic field, described by the
Lagrangian%
\begin{equation}
L=-\frac{1}{4}F_{\mu \nu }^{2}-\partial _{\mu }C^{+}\partial ^{\mu }C+\frac{%
\alpha }{2}\left( \partial _{\mu }A^{\mu }\right) ^{2}\,,  \label{5.2}
\end{equation}%
identity (\ref{3.16}) for the temperature propagator of the field $A_{\mu }$
takes the form%
\begin{equation}
\alpha \partial ^{\mu }\left\langle A_{\mu }A_{\nu }\right\rangle =-\partial
_{\nu }\left\langle CC^{+}\right\rangle \,,  \label{5.3}
\end{equation}%
where we have introduced the notation%
\begin{equation}
\left\langle Q\right\rangle =\mathrm{Sp\,}e^{-\beta H}T_{\tau }Q  \label{5.4}
\end{equation}%
for any operator depending on the ``temperature time'' $\tau $ \cite{10}.
The right- and left-hand sides of relation (\ref{5.3}) can be computed
directly, and, in both cases, they are given by the kernel $\partial _{\nu }%
\frac{1}{\square }$. Nevertheless, the Fourier-image of the l.h.s is known
to contain only even-valued frequencies, whereas the Fourier-image of the
r.h.s., just as Green's functions of any Fermionic operators, contains only
odd-valued frequencies (see also Appendix D). Therefore, relation (\ref{5.3}%
) does not take place.

Since the operator structure and diagrammatic technique for statistical
temperature Green's functions are known \cite{10} to be completely analogous
to the corresponding expressions of field theory, the derivation of the Ward
identities and gauge properties for temperature functions should be carried
out in complete analogy with the corresponding calculations of Section 4.

There is, however, an operation that should be examined in more detail. This
is integration by parts, which has been used several times in Section 4 (see
below). Integration by parts is valid in case the corresponding vertex
conserves momentum (i.e., the sum of the momenta of all the fields in a given
vertex equals to zero). \emph{Within the temperature techniques, this
condition holds true only in case the sum of frequencies of all the fields
in a vertex is even-valued} \cite{10}. As regards the problem we discuss
here, the ``criminal'' cases of integration by parts arise as one passes
from (\ref{3.12}) to (\ref{3.13}), that is, as one shifts the action of a
derivative from the field $t$ to the field $C$ in the l.h.s. of (\ref{3.12}%
), and also as one proves the fact that the second term in the r.h.s. of (%
\ref{3.12}) is equal to zero. Given this, one encounters vertices of the kind%
\begin{equation}
\int dy\,\partial t\,C\,,\;\int dy\,C^{+}CC\,,\;\int dy\,C^{+}CA\,,\;\mathrm{%
etc\,}.  \label{5.5}
\end{equation}

One can observe that integration by parts is valid \emph{in case the
fictitious Fermionic field }$C$\emph{\ (as well as the field }$A$\emph{)
contains only even-valued frequencies}.

Thus, the partition function defined by formula (\ref{5.4}) is not
gauge-invariant. In order to provide gauge-invariance, it is also necessary
that the Green function of a fictitious Fermionic field should contain,
nevertheless, only even-valued frequencies. Besides, the Wick theorem must
also be valid, thus making it possible to deduce relation (\ref{4.3}); see
also Appendix C. Each of these conditions can be fulfilled.

Let us recall that the operator%
\begin{eqnarray}
&&S=e^{-\beta H+\beta \mu _{i}N_{i}}T_{\tau }\exp \left( Q\right) \,,
\label{5.6} \\
&&Q=\int_{0}^{\beta }d\tau \int d^{3}x\left( J^{a,\mu }A_{\mu
}^{a}+J_{i}\varphi _{i}+\overline{\eta }\psi +\overline{\psi }\eta \right)
\label{5.7}
\end{eqnarray}%
can be presented in the form%
\begin{equation}
S=e^{-\beta H+\beta \mu _{i}N_{i}}T_{\tau }\exp \left( Q^{\left( 0\right)
}-\int_{0}^{\beta }d\tau \int d^{3}x\left( H_{\mathrm{int}}^{\left( 0\right)
}\right) \right) \,.  \label{5.8}
\end{equation}%
The dependence of the Heisenberg operators in (\ref{5.6}) on the temperature
parameter $\tau $ is given by the equation%
\begin{equation}
\frac{\partial }{\partial \tau }A_{\mu }^{a}=\left[ \left( H-\mu
_{i}N_{i}\right) ,A_{\mu }^{a}\right] \,.  \label{5.9}
\end{equation}%
For any other operators, it is determined in a similar way and is given by a
formal replacement $it\rightarrow \tau $. In (\ref{5.8}), each of the
operators is free, and its dependence on $\tau $ is determined by the
equation%
\begin{equation}
\frac{\partial }{\partial \tau }A_{\mu }^{a}=\left[ \left( H_{0}-\mu
_{i}N_{i}\right) ,A_{\mu }^{a}\right] \,.  \label{5.10}
\end{equation}%
We have also introduced some terms with chemical potentials (in case they
are necessary), assuming that $N_{i}$ are gauge-invariant.

Consider some matrix element $S$ (which will be denoted by $\left\langle
\;\;\;\right\rangle $). It can be presented in the form%
\begin{equation}
\left\langle S\right\rangle =\left. \exp \left( -H_{\mathrm{int}}\right)
\left\langle S\right\rangle _{0}\right| _{\bar{J}_{\mu }^{a}=\bar{J}%
_{i}=\theta ^{a}=\theta ^{+a}=\bar{\theta}^{a}=\bar{\theta}^{+a}=0}\,.
\label{5.11}
\end{equation}%
In (\ref{5.11}), it is implied that the operator acting on $\left\langle
S\right\rangle _{0}$ is subject to the replacement%
\begin{equation}
A_{\mu }^{a}\rightarrow \frac{\delta }{\delta J_{\mu }^{a}}\,,\;\mathrm{etc}%
\,,  \label{5.12}
\end{equation}%
and the function $\left\langle S\right\rangle _{0}$ is defined as follows:%
\begin{eqnarray}
&&\hspace{-0.5cm}\left\langle S\right\rangle _{0}=\left\langle e^{-\beta
H_{0}+\beta \mu _{i}N_{i}}T_{\tau }\exp \left( Q^{\left( 0\right) }+%
\overline{Q}^{\left( 0\right) }\right) \right\rangle \,,  \label{5.13} \\
&&\hspace{-0.5cm}\overline{Q}=\int_{0}^{\beta }d\tau \int d^{3}x\left(
\overline{J}_{\mu }^{a}\pi ^{a,\mu }+\overline{J}_{i}\Pi _{i}+\theta
^{+a}C^{a}+C^{+a}\theta ^{a}+\overline{\theta }^{+a}\Pi _{C^{+}}^{a}+\Pi
_{C}^{a}\overline{\theta }^{a}\right) \,.  \label{5.14}
\end{eqnarray}%
All the operators in (\ref{5.13}) are free. The matrix elements (\ref{5.11}%
), (\ref{5.13}) obey the Wick theorem provided that the free Green functions
determined by (\ref{5.13}) [i.e., the coefficients of the series expansion
of (\ref{5.13}) in the powers of sources] should decompose into a product of
two-point Green's functions. In other words, expression (\ref{5.13}) as a
function of sources should be given by an exponential of a quadratic
form with respect to the sources. The following matrix element possesses the
required property:%
\begin{equation}
\left\langle S\right\rangle _{0}=\sum_{n}\prod_{i}\lambda
_{i}^{n_{i}}\left\langle n\left| e^{-\beta H_{0}+\beta \mu _{i}N_{i}}T_{\tau
}\exp \left( Q^{\left( 0\right) }+\overline{Q}^{\left( 0\right) }\right)
\right| n\right\rangle \,,  \label{5.15}
\end{equation}%
where%
\begin{equation}
\left| n\right\rangle =\left| n_{1}\right\rangle \otimes \left|
n_{2}\right\rangle \otimes \ldots \,.  \label{5.16}
\end{equation}

$\left\vert n_{i}\right\rangle $ are $n$-particle states, normalized by $\pm
1$, of the particles of $i$-th type, being eigenstates of the free Hamiltonian $%
H_{0}$ and of the Hamiltonian $H_{0}-\mu _{i}N_{i}$; the numbers $%
\lambda _{i}$ depend on the type of a particle (in principle, they may also
depend on the momentum of a particle). In Appendix D, it is shown that (\ref%
{5.15}) equals to%
\begin{equation}
\left\langle S\right\rangle _{0}=Z_{0}^{\left( \lambda _{i}\right) }\exp
\left( \frac{1}{2}J^{\alpha }D_{\alpha \beta }^{H}J^{\beta }\right) \exp
\left( \xi _{i}J_{i}\right) \,,  \label{5.17}
\end{equation}%
where%
\begin{equation}
Z_{0}^{\left( \lambda _{i}\right) }=\left. \left\langle S\right\rangle
_{0}\right\vert _{J^{\alpha }=0}\,,  \label{5.18}
\end{equation}%
while $J^{\alpha }$ denotes the set of all the sources:%
\begin{equation}
J^{\alpha }=\left\{ \overline{J}_{\mu }^{a}\,,\,J^{a,\mu }\,,\,\overline{J}%
_{i}\,,\,J_{i}\,,\,\eta \,,\,\overline{\eta }\,,\,\overline{\theta }%
^{a}\,,\,\theta ^{a}\,,\,\overline{\theta }^{+a}\,,\,\theta ^{+a}\right\} \,,
\label{5.19}
\end{equation}%
and $D_{\alpha \beta }^{H}$ is identical with the two-point Green function
of free fields:%
\begin{equation}
D_{\alpha \beta }^{H}\left( x_{1},x_{2}\right) =\frac{1}{Z_{0}^{\left(
\lambda _{i}\right) }}\sum_{n}\prod_{i}\lambda _{i}^{n_{i}}\left\langle
n\left\vert e^{-\beta H_{0}+\beta \mu _{i}N_{i}}T_{\tau }\omega _{\alpha
}^{\left( 0\right) }\left( x_{1}\right) \omega _{\beta }^{\left( 0\right)
}\left( x_{2}\right) \right\vert n\right\rangle \,.  \label{5.20}
\end{equation}%
$\omega _{\alpha }$ is the set of all the field operators:%
\begin{equation}
\omega _{\alpha }=\left\{ A_{\mu }^{a}\,,\,\pi ^{a,\mu }\,,\,\sigma
_{i}\,,\,\Pi _{i}\,,\,\psi \,,\,\overline{\psi }\,,\,C^{a}\,,\,\Pi
_{C}^{a}\,,\,C^{+a}\,,\,\Pi _{C_{+}}^{a}\right\} \,.  \label{5.21}
\end{equation}%
$x$ is the set of the coordinates: $x=\left( \tau ,\vec{x}\right) $.

Using the equations of motion and equal-time commutation relations for $%
\omega _{\alpha }^{\left( 0\right) }$, one can easily verify that $D_{\alpha
\beta }^{H}$ has the form%
\begin{equation}
\Lambda _{\alpha \beta }^{H}D_{\beta \gamma }^{H}=-\delta _{\alpha \gamma
}\,.  \label{5.22}
\end{equation}%
See the definition of $\Lambda _{\alpha \beta }^{H}$ in Appendix C. Thus,
the quantity%
\begin{equation}
M^{\left( \lambda _{i}\right) }=\frac{1}{Z_{0}^{\left( \lambda _{i}\right) }}%
\left\langle S\right\rangle  \label{5.23}
\end{equation}%
has a typical structure of a generating functional in quantum field theory;
namely, its calculation can be carried out by the same diagrammatic
technique, and it obeys the same functional equations satisfied by a
generating functional of quantum field theory. In particular, a literal
repetition of the reasonings that are presented in quantum field theory
shows \cite{9} that $M^{\left( \lambda _{i}\right) }$ obeys the relation%
\begin{equation}
M^{\left( \lambda _{i}\right) }=e^{\frac{1}{2}\delta \left( 0\right) \mathrm{%
Sp}\ln \alpha _{ab}}M_{W}^{\left( \lambda _{i}\right) }\,,  \label{5.24}
\end{equation}%
where $M_{W}^{\left( \lambda _{i}\right) }$ is computed by the
\textquotedblleft Wick\textquotedblright\ rules%
\begin{equation}
M_{W}^{\left( \lambda _{i}\right) }=\exp \left( L_{\mathrm{int}}\right) \exp
\left( \frac{1}{2}J^{\alpha }D_{\alpha \beta }^{L}J^{\beta }\right) \exp
\left( \xi _{i}J_{i}\right) \,.  \label{5.25}
\end{equation}%
In (\ref{5.25}), the source $J^{\alpha }$ is introduced only for the fields
(that is, $\overline{J}_{\mu }^{a}=\overline{J}_{i}=\overline{\theta }^{a}=%
\overline{\theta }^{+a}=0$), while the Green function of the fields
satisfies the relation%
\begin{equation}
\Lambda _{\alpha \beta }^{L}D_{\beta \gamma }^{L}=-\delta _{\alpha \gamma }
\label{5.26}
\end{equation}%
(the definition of $\Lambda _{\alpha \beta }^{L}$ is given in Appendix C).
Besides, in the course of taking a variation of gauge parameters, $M^{\left(
\lambda _{i}\right) }$ changes as follows:%
\begin{eqnarray}
&&\delta M^{\left( \lambda _{i}\right) }=-e^{-H_{\mathrm{int}%
}}\int_{0}^{\beta }d\tau \int d^{3}x\delta H\exp \left( \frac{1}{2}J^{\alpha
}D_{\alpha \beta }^{H}J^{\beta }\right) e^{\xi _{i}Ji}=  \label{5.27} \\
&&\ =-\frac{1}{Z_{0}^{\left( \lambda _{i}\right) }}\left\langle e^{-\beta
H_{0}+\beta \mu _{i}N_{i}}T_{\tau }\int_{0}^{\beta }d\tau \int d^{3}x\delta
H\exp \left( Q\right) \right\rangle \,,  \label{5.28} \\
&&\delta M_{W}^{\left( \lambda _{i}\right) }=e^{L_{\mathrm{int}%
}}\int_{0}^{\beta }d\tau \int d^{3}x\,\delta L\exp \left( \frac{1}{2}%
J^{\alpha }D_{\alpha \beta }^{L}J^{\beta }\right) e^{\xi _{i}Ji}
\label{5.29}
\end{eqnarray}%
(the sources in (\ref{5.29}) are introduced only for the fields). Relations (%
\ref{5.28}) and (\ref{5.29}) are exact analogues of the corresponding
relations (\ref{A.3.9}) and (\ref{A.3.5}) of quantum field theory. The only
difference between $M^{\left( \lambda _{i}\right) }$ and a generating
functional of quantum field theory arises if one defines the propagators $%
D^{\left( H,L\right) }$ as the integral operators $\Lambda ^{\left(
H,L\right) -1}$, whose unique definition requires that one should impose
certain boundary conditions. The propagators for arbitrary $\lambda _{i}$
are computed in Appendix D.

As has been mentioned in the beginning of this section, it is necessary (in
order to ensure the possibility of deducing the Ward identities) that the
fictitious fields should have only even-valued frequencies. For the
remaining fields, we assume the usual relation between statistics and the
parity of frequencies. As shown in Appendix D, for all the particles, except
the fictitious ones, $\lambda _{i}$ must be chosen as follows:%
\begin{eqnarray}
\lambda_i&=&1\;\mathrm{for\;particles\;with\;positive\;metric}\,,  \notag \\
&&  \label{5.30} \\
\lambda_{i}&=&-1\;\mathrm{for\;particles\;with\;negative\;metric}\,.  \notag
\end{eqnarray}%
Notice that such a choice of $\lambda _{i}$ ensures the coincidence of a
matrix element $\left\langle \ldots \right\rangle $ with the definition of
the trace of an operator (in the subspace of the mentioned particles), so
that the relation between the parity of frequencies and the statistics of a
field can be found without a manifest calculation of the propagator. For the
fictitious particles, the parameters $\lambda _{i}$ must be chosen as
follows:%
\begin{eqnarray}
\lambda_{i}&=&-1\;\mathrm{for\;particles\;with\;positive\;metric}\,,  \notag
\\
&&  \label{5.31} \\
\lambda _{i} &=&1\;\mathrm{for\;particles\;with\;negative\;metric}\,.  \notag
\end{eqnarray}

The series expansion of a fictitious field in the powers of creation and
annihilation operators is made in Appendix A, where it has been shown that
the field $C^{a}$ contains two particles. Correspondingly, $\lambda _{i}$
are equal to%
\begin{equation}
\lambda _{d}=\frac{\varkappa ^{00}}{\left| \varkappa ^{00}\right| }%
\,,\;\lambda _{b}=-\frac{\varkappa ^{00}}{\left| \varkappa ^{00}\right| }\,.
\label{5.32}
\end{equation}%
Notice, once again, that the choice of $\lambda _{i}$ for the fictitious
particles in accordance with (\ref{5.30}) implies that their Green function
contains only odd-valued frequencies. This is a consequence of the fact that
the choice of $\lambda _{i}$ according to (\ref{5.30}) implies that a
matrix element $\left\langle \ldots \right\rangle $ coincides with the trace
of an operator, whereas in this case the propagator of any Fermionic field
contains only odd-valued frequencies.

Consequently, given the choice of $\lambda _{i}$ in accordance with (\ref%
{5.30}), (\ref{5.32}), a literal repetition\footnote{%
Within the temperature techniques, there remain valid the commutation
relations and definitions of canonical momenta (\ref{2.13})--(\ref{2.18})
under the replacement $\partial _{t}\rightarrow i\partial _{\tau }$. This
replacement also implies the coincidence of the equations of motion (\ref%
{2.20}) and (\ref{5.9}), with obvious modifications due to the presence of
the operators $N_{i}$.} of the reasonings of Section 3 makes it possible to
conclude that $M^{\left( \lambda _{i}\right) }$ obeys the following Ward
identity:%
\begin{eqnarray}
&&\,\alpha _{ab}t^{b}\left( x\right) M^{\left( \lambda _{i}\right) }=\frac{1%
}{Z_{0}^{\left( \lambda _{i}\right) }}\left\langle e^{-\beta H_{0}+\beta \mu
_{i}N_{i}}T_{\tau }\int dyQ_{R}^{b}\left( y\right) C^{b}\left( y\right)
C^{+a}\left( x\right) \exp \left( Q\right) \right\rangle =  \notag \\
&&\,=i\int dyQ_{R}^{b}\left( y\right) D^{ba}\left( y,x\right) M^{\left(
\lambda _{i}\right) }\,,  \label{5.33}
\end{eqnarray}%
where $\int dy$ denotes%
\begin{equation}
\int dy\equiv \int_{0}^{\beta }d\tau _{y}\int d^{3}y\,.  \label{5.40}
\end{equation}%
Besides, one ought to remember that in terms of the variables $\tau $ the $D$%
-function is defined by the equation%
\begin{equation}
T^{ab}\left( x\right) D^{bc}\left( x,y\right) =i\delta \left( \tau _{x}-\tau
_{y}\right) \delta \left( \vec{x}-\vec{y}\right)  \label{5.41}
\end{equation}%
and there holds a relation of the form (\ref{3.17}).

A\ variation of the gauge parameters changes $M^{\left( \lambda _{i}\right)
} $ as follows:%
\begin{equation}
\delta M^{\left( \lambda _{i}\right) }=i\int dx\,dy\,Q_{R}^{a}\left(
x\right) D^{ab}\left( x,y\right) \overline{\Lambda }^{b}\left( y\right)
M^{\left( \lambda _{i}\right) }\,,  \label{5.42}
\end{equation}%
where $\overline{\Lambda }$ is given by (\ref{4.8}).

Thus, the statistical average of gauge-invariant operators, and, in
particular, the partition function, are gauge-independent on condition that $%
Z_{0}^{\left( \lambda _{i}\right) }$ should be gauge-invariant.

To prove the gauge invariance of $Z_{0}^{\left( \lambda _{i}\right) }$, let
us apply the formula%
\begin{eqnarray}
&&\delta e^{-\beta \overline{H}}=-e^{-\beta \overline{H}}\int_{0}^{\beta
}d\tau \,e^{\tau \overline{H}}\delta \overline{H}e^{-\tau \overline{H}%
}\equiv e^{-\beta \overline{H}}T_{\tau }\int_{0}^{\beta }d\tau \,\delta
H\left( \tau \right) \,,  \label{5.43} \\
&&\overline{H}=H_{0}-\mu _{i}N_{i}\,.  \label{5.44}
\end{eqnarray}

Thus, the variation of $\ln Z_{0}^{\left( \lambda _{i}\right) }$ equals to
the variation of $M^{\left( \lambda _{i}\right) }$ for the free theory (let
us denote the latter by $M_{0}^{\left( \lambda _{i}\right) }$) with the
vanishing sources:%
\begin{equation}
\delta \ln Z_{0}^{\left( \lambda _{i}\right) }=\left. \delta M_{0}^{\left(
\lambda _{i}\right) }\right\vert _{J^{\alpha }=0}\,.  \label{5.45}
\end{equation}%
From (\ref{5.42}) it follows that $\ln Z_{0}^{\left( \lambda _{i}\right) }$
is gauge-invariant. The gauge invariance of $\ln Z_{0}^{\left( \lambda
_{i}\right) }$ can also be established by a direct calculation using
the expression for $H_{0}$ obtained in Appendices A and E.

In physical gauges (the Coulomb gauge $\varkappa ^{0\mu }=\varkappa
_{i}^{a}=0$, $\varkappa ^{ik}=\delta _{ik}$, $\alpha _{ab}\rightarrow \infty
$; the unitary gauge $\varkappa _{i}^{a}\rightarrow \infty $, that
corresponds to the gauge $B_{m}=0$, with $B_{m}=\xi _{i}^{m}\varphi _{i}$,
being Goldstone Bosons; see, e.g., \cite{11}), non-physical particles are
absent, so that the proposed definition of the partition function and
statistical average, given by%
\begin{equation}
Z=\sum_{n}\prod_{i}\lambda _{i}^{n_{i}}\left\langle n\left| e^{-\beta
H+\beta \mu _{i}N_{i}}\left( \ldots \right) \right| n\right\rangle \,,
\label{5.46}
\end{equation}%
where $\left| n\right\rangle $ and $\lambda _{i}$ are defined in accordance
with (\ref{5.16}), (\ref{5.32}), while $\left( \ldots \right) $ stands for
the $T_{\tau }$-product of operators whose statistical average is to be
found, coincides with the usual definition in physical gauges. Since the
partition function and statistical average of a gauge-invariant operator
are gauge-independent, definition (\ref{5.46}) yields a correct result in
any gauge.

\subparagraph{Acknowledgement}

The author is grateful to A. Linde and E.S. Fradkin for useful discussions.

\appendix

\section{Appendix}

\setcounter{equation}{0}\renewcommand{\theequation}{A.\arabic{equation}}

We are now going to deduce the expressions for the canonical momenta
conjugate to $C$ and $C^{+}$, as well as for their commutation relations
with the help of Schwinger's action principle \cite{6}. Let us present
the part of the action that corresponds to the fictitious fields:%
\begin{equation}
W=\int_{t_{1}}^{t_{2}}dx\left( -\partial _{\mu }C^{+a}\varkappa ^{\mu \nu
}\nabla _{\nu }^{ab}+igC^{+a}\varkappa _{i}^{a}\Gamma _{ij}^{b}\varphi
_{j}C^{b}\right) \,.  \label{A.1.1}
\end{equation}%
We now find a variation $\delta W$:%
\begin{equation}
W=\left. \int_{t_{1}}^{t_{2}}dx\left( \delta C^{+a}\overrightarrow{T}%
^{ab}C^{b}+C^{+a}\overleftarrow{T}^{ab}\Gamma _{ij}^{b}\varphi _{j}\delta
C^{b}\right) +G\right| _{t_{a}}^{t_{2}}\,,  \label{A.1.2}
\end{equation}%
where%
\begin{equation}
G=\int d^{3}x\left( -\delta C^{+a}\varkappa ^{0\nu }\nabla _{\nu
}^{ab}C^{b}-\varkappa ^{0\nu }\partial _{\nu }C^{+a}\delta C^{a}\right) \,.
\label{A.1.3}
\end{equation}%
In accordance with Schwinger's action principle, $G$ is the generator of
variations of the field variables at a fixed moment of time:%
\begin{equation}
\delta C^{a}=i\left[ G,C^{a}\right] \,,\;\delta C^{+a}=i\left[ G,C^{+a}%
\right] \,.  \label{A.1.4}
\end{equation}%
Then, introducing notation (\ref{2.16}), we deduce from (\ref{A.1.3}) and (%
\ref{A.1.4}) that the anticommutation relations (\ref{2.18}) hold true,
whereas the anticommutators between $\Pi _{C}$ and $\Pi _{C^{+}}$, $C$ and $%
C^{+}$ are equal to zero.

Let us now prove the fact that definition (\ref{2.16}) is, in a certain
sense, unique. Namely, we define the canonical momenta as%
\begin{equation}
\Pi _{C}^{a}=-\alpha \varkappa ^{0\mu }\partial _{\mu }C^{+a}\,,\;\Pi
_{C^{+}}^{a}=\beta \varkappa ^{0\mu }\partial _{\mu }C^{b}  \label{A.1.5}
\end{equation}%
and suppose that (\ref{2.18}) holds true. We then construct the Hamiltonian%
\begin{eqnarray}
&&\,H=\alpha _{1}\Pi _{C}\dot{C}+\beta _{1}\Pi _{C}\dot{C}^{+}-L=\frac{1}{%
\varkappa ^{00}}\left( \frac{\alpha _{1}}{\beta }+\frac{\beta _{1}}{\alpha }-%
\frac{1}{\alpha \beta }\right) \Pi _{C}^{a}\Pi _{C^{+}}^{b}-  \notag \\
&&\,-\frac{\alpha _{1}}{\varkappa ^{00}}\Pi _{C}^{a}\left( g\varkappa
^{00}f^{abd}A_{0}^{b}+\varkappa ^{0k}\nabla _{k}^{ad}\right) C^{d}+\partial
_{k}C^{+a}\left( \varkappa ^{ik}-\frac{\varkappa ^{0i}\varkappa ^{0k}}{%
\varkappa ^{00}}\right) \nabla _{i}^{ab}C^{b}-  \notag \\
&&\,-\frac{\beta _{1}}{\varkappa ^{00}}\Pi _{C^{+}}^{a}\varkappa
^{0k}\partial _{k}C^{+a}-igC^{+a}\varkappa _{i}^{a}\Gamma _{ij}^{b}\varphi
_{j}C^{b}\,.  \label{A.1.6}
\end{eqnarray}%
Let us now demand that the Hamiltonian equations of motion in the form (\ref%
{2.20}) for the fields and momenta should yield equations (\ref{2.24}) and
definitions (\ref{A.1.5}). We have%
\begin{equation}
\dot{C}^{a}=\frac{1}{\varkappa ^{00}}\left[ \left( \frac{\alpha _{1}}{\beta }%
+\frac{\beta _{1}}{\alpha }-\frac{1}{\alpha \beta }\right) \Pi
_{C^{+}}^{a}-\alpha _{1}\left( g\varkappa ^{00}\hat{A}_{0}^{ab}+\varkappa
^{0k}\nabla _{k}^{ab}\right) C^{b}\right] \,.  \label{A.1.7}
\end{equation}%
By comparison with (\ref{A.1.5}), we find%
\begin{equation}
\alpha _{1}=1\,,\;\beta _{1}=\frac{1}{\beta }\,.  \label{A.1.8}
\end{equation}%
Next,%
\begin{equation}
\dot{C}^{+a}=-\frac{1}{\beta \varkappa ^{00}}\Pi _{C}^{a}-\frac{1}{\beta
\varkappa ^{00}}\varkappa ^{0k}\partial _{k}C^{+a}\,,  \label{A.1.9}
\end{equation}%
whence%
\begin{equation}
\beta =1=\beta _{1}\,,\;\alpha =\beta \,.  \label{A.1.10}
\end{equation}%
Therefore, all the parameters are defined uniquely, and the expressions
for $\Pi $ and $H$ coincide with the corresponding expressions of Section 2.
Let us now verify that the Lagrangian equations are also fulfilled.

Indeed,%
\begin{eqnarray}
&&\hspace{-1.5cm}\dot{\Pi}_{C}=-\frac{1}{\varkappa ^{_{00}}}\left( g\hat{A}%
^{ab}+\varkappa ^{0k}\nabla _{k}^{ab}\right) \Pi _{C}^{b}+\nabla
_{i}^{ab}\left( \varkappa ^{ik}-\frac{\varkappa ^{0i}\varkappa ^{0k}}{%
\varkappa ^{00}}\right) \partial _{k}C^{+b}+igC^{+b}\varkappa _{i}^{b}\Gamma
_{ij}^{a}\varphi _{j}\,,  \label{A.1.11} \\
&&\hspace{-1.5cm}\dot{\Pi}_{C}=-\frac{1}{\varkappa ^{_{00}}}\varkappa
^{0k}\partial _{k}\Pi _{C}^{a}-\partial _{k}\left( \varkappa ^{ki}-\frac{%
\varkappa ^{0k}\varkappa ^{0i}}{\varkappa ^{00}}\right) \nabla
_{i}^{ab}C^{b}-ig\varkappa _{i}^{a}\Gamma _{ij}^{b}\varphi _{j}C^{b}\,.
\label{A.1.12}
\end{eqnarray}%
Substituting expressions (\ref{2.16}) into (\ref{A.1.11}), (\ref{A.1.12}),
we find the equations of motion (\ref{2.24}).

Let us now expand the operator $C$ in the powers of the creation and
annihilation operators in the free case:%
\begin{eqnarray}
&&L_{0}=-\partial _{\mu }C^{+a}\varkappa ^{\mu \nu }\partial _{\nu
}C^{+a}+C^{+a}m_{ab}C^{b}\,,  \label{A.1.13} \\
&&m_{ab}=-g\varkappa _{i}^{a}\xi _{i}^{b}\,.  \label{A.1.14}
\end{eqnarray}%
For the sake of simplicity, we suppose that there exists a matrix $S$ such
that transforms the matrix $m$ to a diagonal form:%
\begin{equation}
S_{ab}^{-1}m_{bc}S_{cd}=\mu _{a}\delta _{ad}\,.  \label{A.1.15}
\end{equation}

Introducing the fields%
\begin{equation}
V_{a}^{+}=C^{+b}S_{ba}\,,\;V_{a}=S_{ab}^{-1}C^{b}\,,  \label{A.1.16}
\end{equation}%
we bring Lagrangian (\ref{A.1.13}), as well as the equations of motion and
commutation relations, to the form%
\begin{eqnarray}
&&\,L_{0}=-\partial _{\mu }V_{a}^{+}\varkappa ^{\mu \nu }\partial _{\nu
}V_{a}+\mu _{a}V_{a}^{+}V_{a}\,,  \label{A.1.17} \\
&&\,\left( \varkappa ^{\mu \nu }\partial _{\mu }\partial _{\nu }+\mu
_{a}\right) V_{a}=\left( \varkappa ^{\mu \nu }\partial _{\mu }\partial _{\nu
}+\mu _{a}\right) V_{a}^{+}=0\,,  \label{A.1.18} \\
&&\,\left\{ \varkappa ^{0\nu }\partial _{\nu }V_{a}\,,\,V_{b}\right\}
=-\left\{ \varkappa ^{0\nu }\partial _{\nu }V_{a}^{+}\,,\,V_{b}\right\}
=i\delta _{ab}\,,  \label{A.1.19}
\end{eqnarray}%
while the remaining anticommutators are equal to zero.

From (\ref{A.1.18}) it follows that the expansion of the field $V_{a}$ has
the from%
\begin{eqnarray}
&&\hspace{-1.5cm}V_{a}\left( x\right) =\int \frac{dp}{\left( 2\pi \right)
^{3/2}\sqrt{\left\vert \varkappa ^{00}\right\vert }\sqrt{\frac{2}{\varkappa
^{00}}\varkappa ^{0\nu }p_{\nu }}}\left[ e^{-ipx}d_{a}\left( p\right) \delta
\left( p_{0}-\omega _{p}^{a}\right) +b_{a}^{+}\left( p\right) e^{ipx}\delta
\left( p_{0}-\Omega _{p}^{a}\right) \right] \,,  \label{A.1.20} \\
&&\hspace{-1.5cm}\omega _{p}^{a}=-\frac{1}{\varkappa ^{00}}\varkappa
^{0k}p_{k}+\left[ \frac{1}{\varkappa ^{00}}\left( \frac{\varkappa
^{0i}\varkappa ^{0k}}{\varkappa ^{00}}-\varkappa ^{ik}\right) p_{i}p_{k}+%
\frac{\mu _{a}}{\varkappa ^{00}}\right] ^{1/2}\,,  \notag \\
&&\hspace{-1.5cm}\Omega _{p}^{a}=\frac{1}{\varkappa ^{00}}\varkappa
^{0k}p_{k}+\left[ \frac{1}{\varkappa ^{00}}\left( \frac{\varkappa
^{0i}\varkappa ^{0k}}{\varkappa ^{00}}-\varkappa ^{ik}\right) p_{i}p_{k}+%
\frac{\mu _{a}}{\varkappa ^{00}}\right] ^{1/2}\,.  \label{A.1.21}
\end{eqnarray}%
Of course, $\varkappa ^{\mu \nu }$ and $\mu _{a}$ must be subject to such
equations that $\omega _{p}^{a}$ and $\Omega _{p}^{a}$ should be real-valued.

With respect to $V_{a}^{+}$, we assume%
\begin{equation}
V_{a}^{+}=\left( V\right) _{a}^{+}\,.  \label{A.1.22}
\end{equation}%
In the general case, $C^{+}$ is not Hermitian-conjugate to $C$. Then, the
canonical commutation relations lead to the following rules for the
operators $d$ and $C$:%
\begin{equation}
-\left\{ d_{a}\left( p\right) ,d_{b}^{+}\left( q\right) \right\} =\left\{
b_{a}\left( p\right) ,d_{b}^{+}\left( q\right) \right\} =\frac{\varkappa
^{00}}{\left\vert \varkappa ^{00}\right\vert }\delta _{ab}\delta \left( \vec{%
p}-\vec{q}\right) \,.  \label{A.1.23}
\end{equation}%
Let us verify, for instance, (\ref{A.1.19}), namely,%
\begin{eqnarray}
&&\,\left. \left\{ \varkappa ^{0\nu }\partial _{\nu }V_{a}^{+}\left(
x\right) \,,\,V_{b}\left( y\right) \right\} \right\vert _{x_{0}=y_{0}}=-%
\frac{i\delta _{ab}}{2\left\vert \varkappa ^{00}\right\vert }\int \frac{%
\varkappa ^{0\nu }p_{\nu }\,dk\,dp}{\left( 2\pi \right) ^{3}\left[ \frac{1}{%
\varkappa ^{00}}\varkappa ^{0\nu }p_{\nu }\frac{1}{\varkappa ^{00}}\varkappa
^{0\mu }p_{\mu }\right] ^{1/2}}\times  \notag \\
&&\times \,\left[ e^{-ipx+iky}\left( -\frac{\varkappa ^{00}}{\left\vert
\varkappa ^{00}\right\vert }\right) \delta \left( p_{0}-\Omega
_{p}^{a}\right) \delta \left( k_{0}-\Omega _{l}^{a}\right) -\right.  \notag
\\
&&\,-\left. \frac{\varkappa ^{00}}{\left\vert \varkappa ^{00}\right\vert }%
e^{ipx-iky}\delta \left( p_{0}-\omega _{p}^{a}\right) \delta \left(
k_{0}-\omega _{k}^{a}\right) \right] \delta \left( \vec{p}-\vec{k}\right)
=i\delta _{ab}\delta \left( \vec{x}-\vec{y}\right) \,.  \label{A.1.24}
\end{eqnarray}%
The expansion of the initial fields has the form%
\begin{eqnarray}
&&C^{a}\left( x\right) =\frac{1}{\sqrt{\left\vert \varkappa ^{00}\right\vert
}}\int \frac{dp}{\left[ \left( 2\pi \right) ^{3}\frac{2}{\varkappa ^{00}}%
\varkappa ^{0\nu }p_{\nu }\right] ^{1/2}}S_{ac}\left[ e^{-ipx}\delta \left(
p_{0}-\omega _{p}^{c}\right) d_{c}\left( p\right) \right. +  \notag \\
&&+\,\left. e^{ipx}\delta \left( p_{0}-\Omega _{p}^{c}\right)
b_{c}^{+}\left( p\right) \right] \,,  \label{A.1.25} \\
&&C^{+a}\left( x\right) =\frac{1}{\sqrt{\left\vert \varkappa
^{00}\right\vert }}\int \frac{dp}{\left[ \left( 2\pi \right) ^{3}\frac{2}{%
\varkappa ^{00}}\varkappa ^{0\nu }p_{\nu }\right] ^{1/2}}[e^{ipx}\delta
\left( p_{0}-\omega _{p}^{c}\right) d_{c}^{+}\left( p\right) +  \notag \\
&&+e^{-ipx}\delta \left( p_{0}-\Omega _{p}^{c}\right) b_{c}\left( p\right)
]S_{ca}^{-1}\,.  \label{A.1.26}
\end{eqnarray}%
Relation (\ref{A.1.23}) shows that the field $C$ contains two kinds of
Fermions; one of them has a positive norm, while the other one has an
indefinite norm. Of course, this fact is in agreement with the theorem
on the relation between spin and statistics.

Let us, finally, present an expression for the Hamiltonian in terms of the
creation and annihilation operators:%
\begin{eqnarray}
&&\,H_{0}=\int d^{3}x\left[ \frac{1}{\varkappa ^{00}}\Pi _{C}^{a}\Pi
_{C^{+}}^{a}-\frac{1}{\varkappa ^{00}}\Pi _{C}^{a}\varkappa ^{0k}\partial
_{k}C^{a}-\frac{1}{\varkappa ^{00}}\Pi _{C^{+}}^{a}\varkappa ^{0k}\partial
_{k}C^{+a}+\right.  \notag \\
&&\,+\,\,\left. \partial _{i}C^{+a}\left( \varkappa ^{ij}-\frac{\varkappa
^{0i}\varkappa ^{0j}}{\varkappa ^{00}}\right) \partial
_{j}C^{a}+C^{+a}m_{ab}C^{b}\right] =  \notag \\
&&\,=\int d^{3}p\left[ -\frac{\varkappa ^{00}}{\left\vert \varkappa
^{00}\right\vert }\omega _{p}^{a}d_{a}^{+}\left( p\right) d_{a}\left(
p\right) +\frac{\varkappa ^{00}}{\left\vert \varkappa ^{00}\right\vert }%
\Omega _{p}^{a}b_{a}^{+}\left( p\right) b_{a}\left( p\right) \right] \,.
\label{A.1.27}
\end{eqnarray}%
The propagator of the field $C$ is given by%
\begin{eqnarray}
&&\hspace{-1.7cm}\left\langle 0\left\vert TC^{a}\left( x\right) C^{+b}\left(
y\right) \right\vert 0\right\rangle =S_{ac}\left\langle 0\left\vert
TV_{c}\left( x\right) V_{d}^{+}\left( y\right) \right\vert 0\right\rangle
S_{db}^{-1}=  \notag \\
&&\,=\int \frac{dp}{\left( 2\pi \right) ^{4}}e^{-ip\left( x-y\right)
}G_{C}^{ab}\left( p\right) \,,  \label{A.1.28} \\
&&\hspace{-1.7cm}G_{C}^{ab}\left( p\right) =S_{ad}\frac{-i}{\varkappa ^{\mu
\nu }p_{\mu }p_{\nu }-\mu _{d}}S_{db}^{-1}=-i\left( \varkappa ^{\mu \nu
}p_{\mu }p_{\nu }\delta _{ab}-m_{ab}\right) ^{-1}\,.  \label{A.1.29}
\end{eqnarray}

\section{Appendix}

\setcounter{equation}{0} \renewcommand{\theequation}{B.\arabic{equation}}

Let us now prove that the second term in the r.h.s. of (\ref{3.12}) equals
to zero.

We use the antisymmetry property for a product of the fields $C^{a}$ at
coincident points, as well as the possibility of freely integrating by
parts, since the additional terms arising due to the non-commutativity of $%
\partial _{0}$ and $T$-product are proportional to $f^{bdf}\delta _{bd}$,
which is equal to zero. We then have the following equalities (we imply
integration over $dy$ yet do not present it explicitly):%
\begin{eqnarray}
&&\,-g\partial _{\mu }\varkappa ^{\mu \nu }\left( \partial _{\nu
}C^{+b}f^{bdf}C^{f}\right) C^{d}=g\varkappa ^{\mu \nu }\partial _{\mu
}C^{+b}f^{bdf}C^{f}\partial _{\nu }C^{d}=  \notag \\
&&\,=\frac{g}{2}\varkappa ^{\mu \nu }\partial _{\mu }C^{+b}f^{bdf}\left(
C^{f}\partial _{\nu }C^{d}-\partial _{\nu }C^{d}\,C^{f}\right) =\frac{g}{2}%
C^{+b}\overleftarrow{\partial }_{\mu }\varkappa ^{\mu \nu }\overleftarrow{%
\partial }_{\nu }f^{bdf}C^{d}C^{f}\,,  \label{A.2.1} \\
&&\,-g^{2}f^{dcb}f^{fbn}A_{\mu }^{c}\varkappa ^{\mu \nu }\partial _{\nu
}C^{+f}C^{n}C^{d}=  \notag \\
&&\,=\frac{g^{2}}{2}\left( f^{dcb}f^{bfn}-f^{ncb}f^{bfd}\right) A_{\mu
}^{c}\varkappa ^{\mu \nu }\partial _{\nu }C^{+f}\,C^{n}C^{d}=  \notag \\
&&\,=-\frac{g^{2}}{2}C^{+l}\overleftarrow{\partial }_{\mu }\varkappa ^{\mu
\nu }f^{lcb}A_{\nu }^{c}f^{bdf}C^{d}C^{f}\,.  \label{A.2.2}
\end{eqnarray}%
Here, we have used the Jacobi identity%
\begin{equation}
f^{dcb}f^{bfn}+f^{ncb}f^{bdf}=-f^{fcb}f^{bnd}\,.  \label{A.2.3}
\end{equation}%
Next,%
\begin{eqnarray}
&&\,g^{2}C^{+b}\varkappa _{i}^{b}\Gamma _{ij}^{f}\Gamma _{jk}^{d}\varphi
_{k}C^{f}C^{d}=\frac{g^{2}}{2}C^{+b}\varkappa ^{b}\left( \Gamma ^{f}\Gamma
^{d}-\Gamma ^{d}\Gamma ^{f}\right) \varphi C^{f}C^{d}=  \notag \\
&&\,=i\frac{g^{2}}{2}C^{+l}\varkappa _{k}^{l}\Gamma _{kj}^{b}\varphi
_{j}f^{bdf}C^{d}C^{f}\,.  \label{A.2.4}
\end{eqnarray}%
In (\ref{A.2.4}), we have also used (\ref{2.3}). Summarizing (\ref{A.2.1}), (%
\ref{A.2.2}) and (\ref{A.2.4}), we find that the second term in the r.h.s.
of (\ref{3.12}) has the form%
\begin{equation}
\frac{1}{2}gC^{+l}\overleftarrow{T}^{lb}f^{bdf}C^{d}C^{f}\,,  \label{A.2.5}
\end{equation}%
which equals to zero owing to the equations (\ref{2.24}) for $C^{+}$.

\section{Appendix}

\setcounter{equation}{0} \renewcommand{\theequation}{C.\arabic{equation}}

Let us now deduce a formula for the variation of $Z_{W}$ corresponding to
a variation of the parameters in the Lagrangian. In the interaction
representation, $Z_{W}$ can be written as follows:%
\begin{equation}
Z_{W}=\left\langle 0\left| T_{W}\exp \left( iQ+iL_{\mathrm{int}}\right)
\right| 0\right\rangle \,.  \label{A.3.1}
\end{equation}%
When taking a variation of the parameters in the Lagrangian, we need to
examine a variation of the vertices and propagators. According to (\ref%
{A.3.1}), the variation of vertices is given by an insertion into the Green
function of a ``field'' $i\int dx\,\delta L_{\mathrm{int}}$.

Since the field propagator equals to%
\begin{equation}
D_{ij}^{L}=i\left( \Lambda _{ij}^{L}\right) ^{-1}\,,  \label{A.3.2}
\end{equation}%
where $\Lambda ^{L}$ is a differential operator in the free Lagrangian,%
\begin{equation}
L_{0}=\frac{1}{2}\varphi _{i}\Lambda _{ij}^{L}\varphi _{j}\,,  \label{A.3.3}
\end{equation}%
its variation reads%
\begin{equation}
\delta D_{ij}^{L}=D_{ik}^{L}\left( i\delta \Lambda _{kl}^{L}\right)
D_{lj}^{L}\,.  \label{A.3.4}
\end{equation}%
From (\ref{A.3.4}) it follows that the variation of propagators is
equivalent to an insertion of the ``field'' $i\int dx\,\delta L_{0}$. Thus,
the total variation of Green's functions with respect to the parameters of
the Lagrangian is given by an insertion of the ``field'' $i\int dx\,\delta L$%
, namely,%
\begin{equation}
\delta Z_{W}=\left\langle 0\left| T_{W}\left( i\int dx\delta L\right) \exp
\left( iQ+iL_{\mathrm{int}}\right) \right| 0\right\rangle \,.  \label{A.3.5}
\end{equation}

If we do not pass to Wick's rules, $Z$ is given by the formula%
\begin{equation}
\delta Z=\left\langle 0\left\vert T_{W}\exp \left( iQ-iH_{\mathrm{int}%
}\right) \right\vert 0\right\rangle \,.  \label{A.3.6}
\end{equation}%
where $T$ denotes the symbol of the usual $T$-ordering. Let us introduce a
unifying field, $\omega _{\mu }=\left( \pi _{i},\varphi _{i}\right) $. Then (%
\ref{A.3.6}) is computed by the usual Feynman rules with vertices defined
by $-iH_{\mathrm{int}}$ and by the propagator%
\begin{equation}
D_{\mu \nu }^{H}=i\left( \Lambda _{\mu \nu }^{H}\right) ^{-1}\,,
\label{A.3.7}
\end{equation}%
with $\Lambda ^{H}$ being a differential operator in the expression%
\begin{eqnarray}
&&\,\pi _{i}\dot{\varphi}_{i}-H_{0}=\frac{1}{2}\omega _{\mu }\Lambda _{\mu
\nu }^{H}\omega _{\nu }\,,  \notag \\
&&\,\Lambda _{\mu \nu }^{H}=\Lambda _{\mu \nu }^{\left( 1\right) }+\Lambda
_{\mu \nu }^{\left( 2\right) }\,,  \label{A.3.8}
\end{eqnarray}%
where $\Lambda ^{\left( 1\right) }$ is determined by $\pi _{i}\dot{\varphi}%
_{i}$ and is independent of any parameters; $\Lambda ^{\left( 2\right) }$ is
determined by the free Hamiltonian. Next, a literal repetition of the
reasonings that have lead us to formula (\ref{A.3.5}) yields%
\begin{equation}
\delta Z=\left\langle 0\left\vert T\left( i\int dx\,\delta H\right) \exp
\left( iQ-H_{\mathrm{int}}\right) \right\vert 0\right\rangle \,.
\label{A.3.9}
\end{equation}%
We also note that a variation of the Lagrangian and that of the Hamiltonian
with respect to the parameters are related by%
\begin{equation}
-\left. \delta H\right\vert _{\pi ,\varphi }=\left. \delta L\right\vert _{%
\dot{\varphi},\varphi }\,,  \label{A.3.10}
\end{equation}%
where the notation%
\begin{equation*}
\left. \delta A\right\vert _{u}
\end{equation*}%
stands for a variation of $A$ with a fixed $u$.

Formulas (\ref{A.3.5}) and (\ref{A.3.9}) have the same appearance. One must,
however, bear in mind that the time derivatives in (\ref{A.3.6}) commute
with the symbol of $T_{W}$-product, whereas in (\ref{A.3.9}) the time
derivative that arises after the substitution $\pi _{i}\sim \dot{\varphi}%
_{i} $ does not commute with the symbol of $T$-product. Of course, the final
results, calculated by formulas (\ref{A.3.5}) and (\ref{A.3.9}), are
identical, with allowance for relation (\ref{4.2}).

\section{Appendix}

\setcounter{equation}{0} \renewcommand{\theequation}{D.\arabic{equation}}

We are now going to compute the partition functions and Green's functions,
as well as to prove Wick's theorem for various free fields with a
generalized definition of statistical average.

Let us examine, first of all, the case of one degree of freedom.

A) Bose system:%
\begin{eqnarray}
&&\overline{H}=H_{0}-\mu N=\alpha \omega a^{+}a\,,\;\left[ a^{+},a\right]
=\alpha \,,\;\alpha ^{2}=1\,,  \label{A.4.1} \\
&&a\left( \tau \right) =e^{-\omega \tau }a\,,\;a^{+}\left( \tau \right)
=e^{\omega \tau }a^{+}\,,  \label{A.4.2}
\end{eqnarray}%
We now define a generating functional,%
\begin{equation}
Z^{\left( 1\right) }\left( J\right) =\sum_{n}\lambda ^{n}\left\langle
n\left| e^{-\beta \overline{H}}T_{\tau }\exp \left\{ \int_{0}^{\beta }d\tau %
\left[ J\left( \tau \right) a^{+}\left( \tau \right) +J^{+}\left( \tau
\right) a\left( \tau \right) \right] \right\} \right| n\right\rangle \,,
\label{A.4.3}
\end{equation}%
where%
\begin{equation}
\left| n\right\rangle =\frac{1}{\sqrt{n!}}\left( a^{+}\right) ^{n}\left|
0\right\rangle \,,\;\left\langle n\right| =\frac{1}{\sqrt{n!}}\left\langle
0\right| a^{n}\,,\;\lambda ^{2}=1\,.  \label{A.4.3'}
\end{equation}%
Let us transform the $T_{\tau }$-product in (\ref{A.4.3}) to the normal
product:%
\begin{eqnarray}
&&\hspace{-1.5cm}T_{\tau }\exp \left\{ \int_{0}^{\beta }d\tau \left( J\left(
\tau \right) a^{+}\left( \tau \right) +J^{+}\left( \tau \right) a\left( \tau
\right) \right) \right\} =\exp \left\{ \int_{0}^{\beta }d\tau _{1}\,d\tau
_{2}J^{+}\left( \tau _{1}\right) \Delta \left( \tau _{1},\tau _{2}\right)
J\left( \tau _{2}\right) \right\} \times  \notag \\
&&\hspace{-1.5cm}\times :\exp \left\{ J_{1}a^{+}+J_{1}^{+}a\right\} :\,,
\label{A.4.4} \\
&&\hspace{-1.5cm}\Delta \left( \tau _{1},\tau _{2}\right) =\alpha \theta
\left( \tau _{1}-\tau _{2}\right) e^{-\omega \left( \tau _{1}-\tau
_{2}\right) }\,,  \label{A.4.5} \\
&&\hspace{-1.5cm}J_{1}=\int_{0}^{\beta }d\tau \,J\left( \tau \right)
e^{\omega \tau }\,,\;J_{1}^{+}=\int_{0}^{\beta }d\tau \,J^{+}\left( \tau
\right) e^{-\omega \tau }\,.  \label{A.4.6}
\end{eqnarray}%
Substituting (\ref{A.4.4}) into (\ref{A.4.3}), we obtain\footnote{%
We remind that $\mathcal{E}_{n}=\pi n/\beta $.}%
\begin{eqnarray}
&&\,Z_{1}^{\left( 1\right) }=e^{J^{+}\Delta J}\sum_{n=0}^{\infty
}\sum_{k=0}^{\infty }\lambda ^{n}e^{-\beta \omega n}\frac{J_{1}^{k}J_{1}^{+k}%
}{n!k!k!}\left\langle n\left| a^{+k}a^{k}\right| n\right\rangle =  \notag \\
&&\,=e^{J^{+}\Delta J}\sum_{k=0}^{\infty }\frac{\left( \alpha
J_{1}J_{1}^{+}\right) ^{k}}{k!k!}\sum_{n=k}^{\infty }A_{n}^{k}\left( \alpha
\lambda e^{-\omega \beta }\right) ^{n}=  \notag \\
&&\,=e^{J^{+}\Delta J}\sum_{k=0}^{\infty }\frac{\left( \alpha
J_{1}J_{1}^{+}\right) ^{k}}{k!k!}x^{k}\frac{d^{k}}{dx^{k}}\left.
\sum_{n=0}^{\infty }x^{n}\right| _{x=\alpha \lambda e^{-\omega \beta }}=
\notag \\
&&\,=\frac{1}{1-x}e^{J^{+}\Delta J}\sum_{k=0}^{\infty }\frac{1}{k!}\frac{%
\left( \alpha xJ_{1}J_{1}^{+}\right) ^{k}}{k!k!}=  \notag \\
&&\,=\frac{1}{1-x}\exp \left\{ \int_{0}^{\beta }d\tau _{1\,}d\tau
_{2\,}J^{+}\left( \tau _{1}\right) D\left( \tau _{1},\tau _{2}\right)
J^{+}\left( \tau _{2}\right) \right\} \,,  \label{A.4.7} \\
&&\,D\left( \tau _{1},\tau _{2}\right) =\alpha \theta \left( \tau _{1}-\tau
_{2}\right) e^{-\omega \left( \tau _{1}-\tau _{2}\right) }+\frac{\lambda
e^{-\omega \beta -\omega \left( \tau _{1}-\tau _{2}\right) }}{1-\alpha
\lambda e^{-\omega \beta }}\equiv \frac{1}{\beta }\sum_{n}e^{-i\varepsilon
_{n}\left( \tau _{1}-\tau _{2}\right) }D_{1}\left( \varepsilon _{n}\right)
\,,  \label{A.4.8} \\
&&\,D_{1}\left( \varepsilon _{n}\right) =\frac{1}{2}\int_{-\beta }^{\beta
}d\tau \,e^{i\varepsilon _{n}\tau }D_{1}\left( \tau \right) =-\frac{\alpha }{%
2}\frac{1+\alpha \lambda \left( -\right) ^{n}}{i\varepsilon _{n}-\omega }\,.
\label{A.4.9}
\end{eqnarray}%
Next, the partition function $Z_{0}^{\left( 1\right) }$ reads%
\begin{equation}
Z_{0}^{\left( 1\right) }\equiv \sum_{n}\lambda ^{n}\left\langle n\left|
e^{-\beta \overline{H}}\right| n\right\rangle =\frac{1}{1-\alpha \lambda
e^{-\omega \beta }}\,.  \label{A.4.10}
\end{equation}%
As a result, the generating functional takes the form%
\begin{equation}
Z^{\left( 1\right) }\left( J\right) =Z_{0}^{\left( 1\right) }\exp \left(
\frac{1}{2}J^{+}D_{1}J\right) \,,  \label{A.4.11}
\end{equation}%
where $D_{1}$ is given by (\ref{A.4.8}), (\ref{A.4.9}) and equals to%
\begin{equation}
D\left( \tau _{1},\tau _{2}\right) =\frac{1}{Z_{0}^{\left( 1\right) }}%
\sum_{n}\lambda ^{n}\left\langle n\left| e^{-\beta \overline{H}}T_{\tau
}a\left( \tau _{1}\right) a^{+}\left( \tau _{2}\right) \right|
n\right\rangle \,.  \label{A.4.12}
\end{equation}%
Consequently, we can see that the proposed extension of statistical
average obeys Wick's theorem (which is also true for arbitrary $\lambda $).
Besides, the choice%
\begin{equation}
\lambda =\alpha  \label{A.4.13}
\end{equation}%
implies that the $D_{1}$-function contains only even frequencies. As has
been observed in Section 5, in this case there holds the relation%
\begin{equation}
\sum_{n}\alpha ^{n}\left\langle n\left| Q_{1}Q_{2}\right| n\right\rangle
=\sum_{n}\alpha ^{n}\left\langle n\left| Q_{2}Q_{1}\right| n\right\rangle \,,
\label{A.4.15}
\end{equation}%
being a consequence of the identity%
\begin{equation}
1=\sum_{n}\left| n\right\rangle \alpha ^{n}\left\langle n\right| \,,
\label{A.4.16}
\end{equation}%
which makes it possible to find the spectrum of frequencies of the function $%
D_{1}$ by using the Bose-properties of the operators. Nevertheless, the
choice $\lambda =-\alpha $ implies that the function $D_{1}$ (the Green
function of Bose operators) contains only \emph{odd }frequencies. It is
interesting that the distribution function in this case is also of the Fermi
character:%
\begin{equation}
N=\frac{1}{Z_{0}^{\left( 1\right) }}\sum_{n}\left( -\alpha \right)
^{n}\left\langle n\left| e^{-\beta H}\alpha a^{+}a\right| n\right\rangle
=\left( e^{\omega \beta }+1\right) ^{-1}\,.  \label{A.4.17}
\end{equation}%
In a similar way, we deduce the expression for the generating functional of
Green's functions for the field $\varphi \left( \tau \right) =\frac{1}{\sqrt{%
2\omega }}\left( e^{-\omega \tau }a+e^{\omega \tau }a^{+}\right) $, namely,%
\begin{eqnarray}
&&\,Z^{\left( 2\right) }\equiv \sum_{n}\lambda ^{n}\left\langle n\left|
e^{-\beta \overline{H}}T_{\tau }\exp \left( \int_{0}^{\beta }d\tau
\,J\varphi \right) \right| n\right\rangle =Z_{0}^{\left( 1\right) }\exp
\left( \frac{1}{2}JD_{2}J\right) \,,  \label{A.4.18} \\
&&\,D_{2}\left( \tau _{1},\tau _{2}\right) =\frac{1}{Z_{0}^{\left( 1\right) }%
}\sum_{n}\lambda ^{n}\left\langle n\left| e^{-\beta \overline{H}}T_{\tau
}\varphi \left( \tau _{1}\right) \varphi \left( \tau _{2}\right) \right|
n\right\rangle =\frac{1}{\beta }\sum_{n}e^{-\varepsilon _{n}\left( \tau
_{1}-\tau _{2}\right) }D_{2}\left( \varepsilon _{n}\right) \,,
\label{A.4.19} \\
&&\,D_{2}\left( \varepsilon _{n}\right) =-\frac{\alpha }{2}\frac{1+\alpha
\lambda \left( -\right) ^{n}}{\left( i\varepsilon _{n}\right) ^{2}-\omega
^{2}}  \label{A.4.20}
\end{eqnarray}

B) Fermi case:%
\begin{eqnarray}
&&\,\overline{H}=\alpha \omega a^{+}a\,,\;\left\{ a,a^{+}\right\} =\alpha
\,,\;\alpha ^{2}=1\,,  \label{A.4.21} \\
&&\,T_{\tau }\exp \left\{ \int_{0}^{\beta }d\tau \,\left( \eta ^{+}\left(
\tau \right) a\left( \tau \right) +a^{+}\left( \tau \right) \eta \left( \tau
\right) \right) \right\} =e^{\eta ^{+}\Delta \eta }:\exp \left( \eta
_{1}^{+}a+a^{+}\eta _{1}\right) :\,,  \label{A.4.22} \\
&&\,\eta _{1}^{+}=\int_{0}^{\beta }d\tau \,e^{-\omega \tau }\eta ^{+}\left(
\tau \right) \,,\;\eta _{1}=\int_{0}^{\beta }d\tau \,e^{\omega \tau }\eta
\left( \tau \right) \,.\,  \label{A.4.23}
\end{eqnarray}%
Calculating the generating functional, one should take into account the
anticommutation properties of $a$ and $\eta $:%
\begin{equation}
a^{2}=a^{+2}=\eta _{1}^{2}=\eta _{1}^{+2}=0\,.  \label{A.4.24}
\end{equation}%
We have%
\begin{eqnarray}
&&\,Z_{0}^{\left( 2\right) }=\sum_{n}\lambda ^{n}\left\langle n\left|
e^{-\beta \overline{H}}\right| n\right\rangle =1+\alpha \lambda e^{-\beta
\omega }\,,  \label{A.4.25} \\
&&\,Z^{\left( 2\right) }\left( \eta \right) =\sum_{n}\lambda
^{n}\left\langle n\left| e^{-\beta \overline{H}}T_{\tau }\exp \left(
\int_{0}^{\beta }d\tau \,\left( \eta ^{+}a+a^{+}\eta \right) \right) \right|
n\right\rangle =  \notag \\
&&\,=e^{\eta ^{+}\Delta \eta }\left[ 1+\alpha \lambda e^{-\beta \omega
}-\lambda \eta _{1}^{+}\eta _{1}e^{-\omega \beta }\left\langle 1\left|
a^{+}a\right| 1\right\rangle \right] =Z_{0}^{\left( 2\right) }\exp \left(
\eta ^{+}D_{3}\eta \right) \,,  \notag \\
&&\,D_{3}\left( \tau _{1},\tau _{2}\right) =\alpha \theta \left( \tau
_{1}-\tau _{2}\right) e^{-\omega \left( \tau _{1}-\tau _{2}\right) }-\frac{%
\lambda e^{-\omega \beta -\omega \left( \tau _{1}-\tau _{2}\right) }}{%
1+\alpha \lambda e^{-\omega \beta }}=  \label{A.4.26} \\
&&\,=\frac{1}{\beta }\sum_{n}e^{-i\varepsilon _{n}\left( \tau _{1}-\tau
_{2}\right) }D_{3}\left( \varepsilon _{n}\right) \,,  \label{A.4.27} \\
&&\,D_{3}\left( \varepsilon _{n}\right) =-\frac{\alpha }{2}\frac{1-\alpha
\lambda \left( -\right) ^{n}}{i\varepsilon _{n}-\omega }\,.  \label{A.4.27'}
\end{eqnarray}%
We can see, once again, that the choice $\lambda =\alpha $ implies that the
Green function, just as it should be in the case of a Fermi Green function,
contains only odd frequencies. However, in the case $\lambda =-\alpha $
a Fermi Green function contains only \emph{even} frequencies. This definition
leads to a Bose distribution function of a Fermion:%
\begin{equation}
N=\frac{1}{Z_{0}^{\left( 2\right) }}\sum_{n}\lambda ^{n}\left\langle n\left|
e^{-\beta \overline{H}}\alpha a^{+}a\right| n\right\rangle =\left( e^{\omega
\beta }-1\right) ^{-1}\,.  \label{A.4.28}
\end{equation}%
Wick's theorem, once again, is valid for an arbitrary $\lambda $. A\
generalization of the above reasoning to the case of a system of particles
in a space of arbitrary dimension is evident.

We finally compute the partition function and Green function of a
fictitious particle, whose field operator is given by the formulas [see,
(\ref{A.1.25}), (\ref{A.1.26}) and (\ref{A.1.27})]
\begin{eqnarray}
&&\,C^{a}\left( x\right) =S_{ab}V_{b}\left( x\right) \,,\;C^{+a}\left(
x\right) =V_{b}^{+}\left( x\right) S_{ba}\,,  \label{A.4.29} \\
&&\,V_{a}\left( x\right) =\int \frac{d^{3}p}{\left( 2\pi \right) ^{3/2}\sqrt{%
2\left| \varkappa ^{00}\right| \nu _{p}^{a}}}\left[ e^{-\omega _{p}^{a}\tau
-i\vec{p}\vec{x}}d_{a}\left( p\right) +e^{\Omega _{p}^{a}\tau +i\vec{p}\vec{x%
}}b_{a}^{+}\left( p\right) \right] \,,  \label{A.4.30} \\
&&\,V_{a}^{+}\left( x\right) =\int \frac{d^{3}p}{\left( 2\pi \right) ^{3/2}%
\sqrt{2\left| \varkappa ^{00}\right| \nu _{p}^{a}}}\left[ e^{\omega
_{p}^{a}\tau +i\vec{p}\vec{x}}d_{a}^{+}\left( p\right) +e^{-\Omega
_{p}^{a}\tau -i\vec{p}\vec{x}}b_{a}\left( p\right) \right] \,,
\label{A.4.31} \\
&&\,\nu _{p}^{a}=\left[ \frac{1}{\varkappa ^{00}}\left( \frac{\varkappa
^{0i}\varkappa ^{0k}}{\varkappa ^{00}}-\varkappa ^{ik}\right) p_{i}p_{k}+%
\frac{1}{\varkappa ^{00}}\mu _{a}\right] ^{1/2}\,.  \label{A.4.32}
\end{eqnarray}%
We restrict ourselves to the following values of $\lambda _{d}$ and $\lambda
_{b}$:%
\begin{equation}
\lambda _{b}=-\lambda _{d}=\lambda \frac{\varkappa ^{00}}{\left| \varkappa
^{00}\right| }\,,\;\lambda ^{2}=1\,.  \label{A.4.33}
\end{equation}%
We find%
\begin{eqnarray}
&&\,Z_{0}=\prod_{a,p}\left( 1+\lambda e^{-\omega _{p}^{a}\beta }\right)
\left( 1+\lambda e^{-\Omega _{p}^{a}\beta }\right) \,,  \label{A.4.34} \\
&&\,Z\left( \theta \right) =\sum_{n}\lambda _{d}^{n_{d}}\lambda
_{b}^{n_{b}}\left\langle n\left| e^{-\beta H}T_{\tau }\exp \left[
\int_{0}^{\beta }d\tau \,\left( \theta ^{+a}C^{a}\left( x\right)
+C^{+a}\left( y\right) \theta ^{a}\right) \right] \right| n\right\rangle =
\label{A.4.35} \\
&&\,=Z_{0}\exp \left( \theta ^{+a}D_{ab}\theta ^{b}\right) \,,  \notag \\
&&\,D_{ab}\left( x-y\right) =\frac{1}{Z_{0}}\sum_{n}\lambda
_{d}^{n_{d}}\lambda _{b}^{n_{b}}\left\langle n\left| e^{-\beta H}T_{\tau
}C^{a}\left( x\right) C^{+b}\left( y\right) \right| n\right\rangle =  \notag
\\
&&\,=S_{ad}\frac{1}{\beta }\sum_{n}\int \frac{d^{3}p}{\left( 2\pi \right)
^{3}}e^{-i\varepsilon _{n}\left( \tau _{x}-\tau _{y}\right) -i\vec{p}\left(
\vec{x}-\vec{y}\right) }D_{d}\left( \varepsilon _{n},\vec{p}\right)
S_{db}^{-1}\,,  \label{A.4.36} \\
&&\,D_{a}\left( \varepsilon _{n},\vec{p}\right) =\frac{1-\left( -\right)
^{n}\lambda }{2\left( \varkappa ^{\mu \nu }p_{\mu }p_{\nu }-\mu _{a}\right) }%
\,.  \label{A.4.37}
\end{eqnarray}%
In (\ref{A.4.37}), one needs to make a replacement:\ $p_{0}\rightarrow
i\varepsilon _{n}$. We can see that the choice $\lambda =+1$ leads to a
Green function (\ref{A.4.37}) that contains only odd frequencies. However,
the case $\lambda =-1$, corresponding to the choice of $\lambda _{d}$ and $%
\lambda _{b}$ indicated in Section 5 [see, (\ref{5.31}), (\ref{5.32})], leads
to a Green function of Fermionic fictitious particles that contains only
even frequencies.

Using the reasonings that have been presented in this appendix, one can
easily see that the generating functional for an arbitrary field linear in
the creation and annihilation operators can be computed with the help of
Wick's theorem:%
\begin{equation}
Z\left( J\right) =\sum_{n}\prod_{i}\lambda _{i}^{n_{i}}\left\langle n\left|
e^{-\beta \overline{H}}T_{\tau }e^{J\omega }\right| n\right\rangle =Z_{0}e^{%
\frac{1}{2}JDJ}\,,  \label{A.4.38}
\end{equation}%
where $D$ is the Green function of the field $\omega $:%
\begin{equation}
D=\frac{1}{Z_{0}}\sum_{n}\prod_{i}\lambda _{i}^{n_{i}}\left\langle n\left|
e^{-\beta \overline{H}}T_{\tau }\omega \omega \right| n\right\rangle \,.
\label{A.4.39}
\end{equation}%
This makes it possible to prove (\ref{5.17}) and (\ref{5.20}). In order to
prove (\ref{5.22}), one has to use the equations of motion and canonical
commutation relations for $\omega $.

\section{Appendix}

\setcounter{equation}{0} \renewcommand{\theequation}{E.\arabic{equation}}

Let us now perform a canonical quantization of the system of free fields $%
A_{\mu }^{a}$, being Goldstone Bosons. In view of a tedious character of the
resulting formulas, we restrict ourselves to the case $\varkappa
_{i}^{a}\sim \xi _{i}^{a}$, $\alpha _{ab}=\alpha \delta _{ab}$, $\varkappa
^{\mu \nu }=g^{\mu \nu }$. Then the system which consists of a multiplet of
vector and scalar fields is equivalent to a set of Abelian vector fields,
each interacting only with its scalar field. The Lagrangian of this system
of fields $\Omega _{\beta }=\left( A_{\mu },\sigma \right) $ has the form%
\begin{equation}
L=-\frac{1}{4}F_{\mu \nu }^{2}+\frac{M^{2}}{2}A_{\mu }^{2}+M\sigma \partial
_{\mu }A_{\mu }+\frac{1}{2}\partial _{\mu }\sigma \partial _{\mu }\sigma +%
\frac{\alpha }{2}\left( \partial _{\mu }A_{\mu }+\beta \sigma \right) ^{2}\,.
\label{A.5.1}
\end{equation}%
Let us present the field $A_{\mu }$ as follows:%
\begin{equation}
A_{\mu }=V_{\mu }+\frac{1}{2}\partial _{\mu }\varphi \,,\;\partial _{\mu
}V_{\mu }=0\,.  \label{A.5.2}
\end{equation}%
Substituting expansion (\ref{A.5.2}) into the equations of motion for $%
A_{\mu }$ and $\sigma $, that follow from (\ref{A.5.1}), we find%
\begin{eqnarray}
&&\,\left( \Box +M^{2}\right) V_{\mu }=0\,,  \label{A.5.3} \\
&&\,\left( \Box -\frac{M^{2}}{\alpha }\right) \varphi +\left( \beta M+\frac{%
M^{2}}{\alpha }\right) \sigma =\left( 1+\frac{\alpha \beta }{M}\right) \Box
\varphi -\left( \Box -\alpha \beta ^{2}\right) \sigma =0\,.  \label{A.5.4}
\end{eqnarray}%
From (\ref{A.5.4}) it follows that%
\begin{equation}
\left( \Box +\beta M\right) ^{2}\varphi =\left( \Box +\beta M\right)
^{2}\sigma =0\,.  \label{A.5.5}
\end{equation}%
Consequently, we can see that $V_{\mu }$ is a usual massive vector particle,
whereas the fields $\varphi $ and $\sigma $ have to be decomposed into
summands whose Fourier transformations are proportional to $\delta \left(
k^{2}-\beta M\right) $ and $\delta ^{\prime }\left( k^{2}-\beta M\right) $.
Substituting this decomposition into (\ref{A.5.4}), we can see that amongst
the $8$ amplitudes there are only $4$ arbitrary ones. In addition, the
fields $\varphi $ and $\sigma $ can be presented in the form%
\begin{eqnarray}
&\!\!\!\!\!\!\!\!\!\!\!\!\varphi=\int\frac{d^4x\sqrt{2k_0}}{(2\pi)^{3/2}}%
\theta \left( k_{0}\right) \left\{ e^{-ikx}\left[ \delta \left( k^{2}-\beta
M\right) c_{k}+\left( \beta M+\frac{M^{2}}{\alpha }\right) \delta ^{\prime
}\left( k^{2}-\beta M\right) \left( c_{k}-d_{k}\right) \right] +\mathrm{h.c.}%
\right\} ,  \label{A.5.6} \\
&\!\!\!\!\!\!\!\!\!\!\!\!\sigma=\int\frac{d^4x\sqrt{2k_0}}{(2\pi)^{3/2}}%
\theta \left( k_{0}\right) \left\{ e^{-ikx}\left[ \delta \left( k^{2}-\beta
M\right) d_{k}+\left( \beta M+\frac{M^{2}}{\alpha }\right) \delta ^{\prime
}\left( k^{2}-\beta M\right) \left( c_{k}-d_{k}\right) \right] +\mathrm{h.c.}%
\right\}.  \label{A.5.7}
\end{eqnarray}%
In quantum field theory, $c_{k}$, $d_{k}$ and $c_{k}^{+}$, $d_{k}^{+}$ are
operators. Their commutators can be found due to the commutation relations
between the fields and conjugate momenta. Let us present a qualitative
analysis of finding these relations. The fact that $\sigma $ and $\Pi
_{\sigma }=\dot{\sigma}$ must commute with $A_{\mu }$ implies that $\sigma $
and $\dot{\sigma}$ commute with $\varphi $, whence it follows that $\left[
c,d^{+}\right] =0$. In order that the commutator function should not contain
$\delta ^{n}$, it is necessary that the equality $\left[ c,c^{+}\right] +%
\left[ d,d^{+}\right] =0$ must take place. Finally, the canonical
commutation relations of $\sigma $ and $\Pi _{\sigma }$ yield $\left[ d,d^{+}%
\right] =1$.

Consequently, we find that the field $\Omega _{\beta }$ is presented in the
form (\ref{A.5.2}), (\ref{A.5.6}), (\ref{A.5.7}), where%
\begin{equation}
V_{\mu }=\int \frac{d^{4}x\sqrt{2k_{0}}}{\left( 2\pi \right) ^{3/2}}\theta
\left( k_{0}\right) \left\{ e^{-ikx}\delta \left( k^{2}-M^{2}\right)
u_{k}^{l}a_{k}^{l}+\mathrm{h.c.}\right\} \,,  \label{A.5.8}
\end{equation}%
$u_{k}^{l}$ being three orthonormalized transversal polarization vectors.
The creation and annihilation operators obey the commutation relations%
\begin{equation}
\left[ a_{k}^{l},a_{p}^{+l^{\prime }}\right] =\delta _{ll^{\prime }}\delta
\left( \vec{k}-\vec{p}\right) \,,\;\left[ d_{k},d_{p}^{+}\right] =-\left[
c_{k},c_{p}^{+}\right] =\delta \left( \vec{k}-\vec{p}\right) \,,
\label{A.5.9}
\end{equation}%
the remaining commutators being equal to zero. By direct calculation, we can
prove that all the canonical commutators between the fields $\Omega _{\beta
} $ and the canonical momenta constructed from Lagrangian (\ref{A.5.1})
are fulfilled.

The Hamiltonian of the system equals to%
\begin{eqnarray}
&&\,H=\int d^{3}k\left[ \sum_{l}\omega _{k}a_{k}^{+l}a_{k}^{l}+\Omega
_{k}d_{k}^{+l}d_{k}^{l}-\Omega _{k}c_{k}^{+}c_{k}\right] +  \notag \\
&&\,+\,\,\frac{\beta M+\frac{M^{2}}{\alpha }}{2\Omega _{k}}\left(
c_{k}^{+}-d_{k}^{+}\right) \left( c_{k}-d_{k}\right) \,,  \label{A.5.10} \\
&&\,\omega _{k}=\sqrt{\vec{k}^{2}+M^{2}}\,,\;\Omega _{k}=\sqrt{\vec{k}%
^{2}+\beta M}\,,  \label{A.5.11}
\end{eqnarray}

The $\delta ^{\prime }$-function can be presented as $\frac{1}{k_{0}}%
\partial /\partial k_{0}\delta \left( k^{2}-\beta M\right) $ and then
integrated by parts. As a result, we find that the fields $\varphi $ and $%
\sigma $ can be presented in the form (\ref{A.5.6}), (\ref{A.5.7}), with the
following replacement:%
\begin{equation}
\delta ^{\prime }\left( k^{2}-\beta M\right) \rightarrow \left( \frac{1}{%
4k_{0}^{2}}+\frac{it}{2k_{0}}\right) \delta \left( k^{2}-\beta M\right) \,.
\label{A.5.12}
\end{equation}%
This representation is useful in the case $\beta =0$, when the $\delta
^{\prime }$-function needs an additional determination.

In the general case, described by the Lagrangian%
\begin{eqnarray}
&&\,L=-\frac{1}{4}F_{\mu \nu }^{a}F^{a,\mu \nu }+\frac{M_{0}^{2}}{2}A_{\mu
}^{a}A^{a,\mu }+M^{2}\sigma ^{a}\partial _{\mu }A^{a,\mu }+\frac{1}{2}%
\partial _{\mu }\sigma ^{a}\partial ^{_{\mu }}\sigma ^{a}+\frac{1}{2}%
t^{a}\alpha _{ab}t^{b}\,,  \notag \\
&&\,t^{a}=\varkappa ^{\mu \nu }\partial _{\mu }A_{\nu }^{a}+\varkappa
_{ab}\sigma ^{b}\,,  \label{A.5.13}
\end{eqnarray}%
one should proceed in a similar way. Let us now introduce a field $V_{\mu
}^{a}$,%
\begin{equation}
V_{\mu }^{a}=A_{\mu }^{a}-\frac{1}{M_{a}}\partial _{\mu }\sigma
^{a}\,,\;t^{a}=\varkappa ^{\mu \nu }\partial _{\mu }V_{\nu }^{a}+\frac{1}{%
M_{a}}\left( \varkappa ^{\mu \nu }\partial _{\mu }\partial _{\nu }\delta
_{ab}+M_{a}\varkappa _{ab}\right) \sigma ^{b}\,.  \label{A.5.14}
\end{equation}%
In terms of this field, the equations of motion acquire the form%
\begin{eqnarray}
&&\,\Lambda _{\mu \nu }^{ab}V^{b,\nu }=0\,,\;M^{a}\partial _{\mu }V^{a,\mu
}+\varkappa _{ba}\alpha _{bc}t^{c}=0\,,  \notag \\
&&\,\Lambda _{\mu \nu }^{ab}=\delta _{ab}\left[ g_{\mu \nu }\left( \Box
+M_{a}^{2}\right) -\partial _{\mu }\partial _{\nu }\right] +\widehat{%
\partial }_{\mu }\partial _{\nu }\varkappa _{ab}^{-1}\,,\;\widehat{\partial }%
_{\mu }=\varkappa _{\mu }^{\,\,\,\nu }\partial _{\nu }\,.  \label{A.5.15}
\end{eqnarray}%
We now have to find the spectrum of the system. For this purpose, we need to
calculate $\det \Lambda $, namely,%
\begin{eqnarray}
&&\,\ln \det \Lambda =\mathrm{Sp\,\ln }\Lambda =\left. \mathrm{Sp\,\ln }%
\lambda ^{-1}\left( 1+\lambda \Lambda \right) \right| _{\lambda =\infty }=
\notag \\
&&\,=\mathrm{Sp\,\ln }\lambda +\sum_{k}\mathrm{Sp}\left( -\right)
^{k}\lambda ^{k}\left[ \left( g_{\mu \nu }-\frac{\partial _{\mu }\partial
_{\nu }}{\Box }\right) \left( \Box +M_{a}^{2}\right) \delta _{ab}\right. +
\notag \\
&&\,+\left. \left( \frac{\partial _{\mu }\partial _{\nu }}{\Box }%
M_{a}^{2}\delta _{ab}+\widehat{\partial }_{\mu }\partial _{\nu
}M_{a}\varkappa _{ab}^{-1}\right) \right] ^{k}=\mathrm{Sp\,\ln }\lambda +%
\frac{3}{4}\mathrm{Sp\,}\delta _{\mu \nu }\delta _{ab}\ln \left[ 1+\lambda
\left( \Box +M_{a}^{2}\right) \right] +  \notag \\
&&\,+\left. \frac{1}{4}\mathrm{Sp\,}\delta _{\mu \nu }\ln \left[ 1+\lambda
M\varkappa ^{-1}\left( \widehat{\partial }_{\sigma }\partial ^{\sigma
}+\varkappa M\right) \right] \right| _{\lambda =\infty }=  \notag \\
&&\,=\det \,^{-1}\varkappa _{ab}\det \,\left( \widehat{\partial }_{\mu
}\partial ^{\mu }+\varkappa _{ab}M_{b}\right) \prod_{a}M_{a}\det
\,^{3}\left( \Box +M_{a}^{2}\right) \,.  \label{A.5.16}
\end{eqnarray}%
Thus, the spectrum of the system is determined by solutions of the equations%
\begin{equation}
\Box +M_{a}^{2}=0\,,\;\widehat{\partial }_{\mu }\partial ^{\mu }+\varkappa
_{ab}\mu _{b}=0\,,  \label{A.5.17}
\end{equation}%
where $\mu _{a}$ stand for the eigenvalues of the matrix $%
m_{ab}=M_{a}\varkappa _{ab}$. Taking account of the expression (\ref{A.5.14}%
) for $t^{a}$, we can see that the field $V_{\mu }^{a}$ has to be decomposed
in $\delta \left( k^{2}-M_{a}^{2}\right) $ and $\delta \left( \varkappa
^{\mu \nu }k_{\mu }k_{\nu }-\mu _{a}\right) $, whereas the field $\sigma $
should be decomposed in $\delta \left( k^{2}-M_{a}^{2}\right) $, $\delta
\left( \varkappa ^{\mu \nu }k_{\mu }k_{\nu }-\mu _{a}\right) $ and $\delta
^{\prime }\left( \varkappa ^{\mu \nu }k_{\mu }k_{\nu }-\mu _{a}\right) $.

Let us also present the expressions for (Wick's) field propagators. The
simplest way to find them is to calculate the Gaussian functional integral
of $L$. In addition, we can see that if one introduces sources for the
fields $V_{\mu }^{a}$ and $t^{a}$ (rather than those for $A_{\mu }^{a}$ and $%
\sigma ^{a}$) the integrals over $V_{\mu }^{a}$ and $t^{a}$ factor out and
can be easily computed. As a result, we have%
\begin{eqnarray}
&&\,\left\langle V_{\mu }^{a}\,V_{\nu }^{a}\right\rangle =i\delta
_{ab}\left( g_{\mu \nu }+\frac{\partial _{\mu }\partial _{\nu }}{M_{a}^{2}}%
\right) \frac{1}{\Box +M_{a}^{2}}\,,  \label{A.5.18} \\
&&\,\left\langle V_{\mu }^{a}\,t^{a}\right\rangle =0\,,  \label{A.5.19} \\
&&\,\left\langle t^{a}\left( x\right) \,t^{a}\left( y\right) \right\rangle
=i\alpha ^{ab}\delta \left( x-y\right) \,.  \label{A.5.20}
\end{eqnarray}%
Hence, the propagators of the fields $A_{\mu }^{a}$ and $\sigma ^{a}$ are
found with the help of the relations%
\begin{eqnarray}
&&\,\sigma ^{a}=\left( \varkappa ^{\mu \nu }\partial _{\mu }\partial _{\nu
}\delta _{ab}+m_{ab}\right) ^{-1}M_{b}\left( t^{b}-\varkappa ^{\mu \nu
}\partial _{\mu }V_{\nu }^{b}\right) \,,  \notag \\
&&\,A_{\mu }^{a}=V_{\mu }^{a}+\frac{1}{M_{a}}\partial _{\mu }\left(
\varkappa ^{\mu \nu }\partial _{\mu }\partial _{\nu }\delta
_{ab}+m_{ab}\right) ^{-1}M_{b}\left( t^{b}-\varkappa ^{\mu \nu }\partial
_{\mu }V_{\nu }^{b}\right) \,.  \label{A.5.21}
\end{eqnarray}

\section{Appendix}

\setcounter{equation}{0} \renewcommand{\theequation}{F.\arabic{equation}}

In this appendix, we deduce, for the sake of completeness, the Ward
identities in case the Green functions include the fictitious fields.
Deriving these identities is quite similar to the corresponding calculation
of Section 3, and, therefore, we are not going to present a detailed
analysis.

Consider the generating functionals%
\begin{eqnarray}
&&\,Z_{C_{y}^{d}C_{x}^{+a}}=\left\langle 0\left| T\exp \left( i\hat{Q}%
\right) c^{d}\left( y\right) c^{+a}\left( y\right) \right| 0\right\rangle \,,
\label{A.6.1} \\
&&\,Z=\left\langle 0\left| T\exp \left( i\hat{Q}\right) \right|
0\right\rangle \,,  \label{A.6.2}
\end{eqnarray}%
where%
\begin{equation}
\hat{Q}=Q+\theta ^{+a}C^{a}+C^{+a}\theta ^{a}\,.  \label{A.6.3}
\end{equation}%
An analogue of formula (\ref{3.12}) now takes the form%
\begin{eqnarray}
&&\hspace{-1.7cm}\int dy\,t^{f}\left( y\right) \alpha
_{fc}T^{cd}Z_{CC^{+}}=\int dy\left\langle \left[ g\theta ^{+b}\left(
y\right) f^{bdn}C^{n}\left( y\right) -Q_{R}^{d}\left( y\right) \right]
C^{d}\left( y\right) C^{+a}\left( y\right) \right\rangle +  \notag \\
&&\hspace{-1.7cm}+\int dy\left\langle \left[ -g\varkappa ^{\mu \nu }\nabla
_{\mu }^{db^{\prime }}\left( \partial _{\nu }C^{+b}\left( y\right)
f^{bb^{\prime }f}C^{f}\left( y\right) \right) +g^{2}C^{+b}\left( y\right)
\varkappa ^{b}\Gamma ^{f}\Gamma ^{d}\varphi \left( y\right) C^{f}\left(
y\right) \right] C^{d}\left( y\right) C^{+a}\left( x\right) \right\rangle \,.
\label{A.6.4}
\end{eqnarray}%
In this appendix, we use the following notation:%
\begin{equation}
\left\langle \ldots \right\rangle =\left\langle T\left( \ldots \right) \exp
\left( i\hat{Q}\right) \right\rangle \,.  \label{A.6.5}
\end{equation}%
As in Section 3, we need to carry out an integration by parts in the second
term of (\ref{A.6.4}). We, however, must take into account that $%
\hat{Q}$ depends on $C$ and $C^{+}$, and, therefore, $\partial _{0}$ does
not commute with the symbol of $T$-product. We need to extract $\partial
_{0} $ from the symbol of $T$-product, then integrate by parts, using the
antisymmetry of the operators $C$ (as has been done in Appendix B; one only
has to remember that $\partial _{0}$ now stands aside from the symbol of $T$%
-product), then integrate by parts once again, and finally insert $%
\partial _{0}$ in the symbol of $T$-product. In transposing $\partial _{0}$
with the symbol of $T$-product, we need formula (\ref{3.3}), while also
making the replacement $Q\rightarrow \hat{Q}$. As a result, the Ward
identities have a surprisingly simple form:%
\begin{eqnarray}
&&\,i\alpha _{ab}t^{b}\left( x\right) Z=-\left\langle \int
dy\,Q_{R}^{b}\left( y\right) C^{b}\left( y\right) C^{+a}\left( x\right)
\right\rangle +\left\langle \int dy\,t^{b}\left( y\right) \alpha _{bd}\theta
^{d}\left( y\right) C^{+a}\left( x\right) \right\rangle -  \notag \\
&&\,-\,\,\frac{g}{2}\left\langle \int dy\,\theta ^{+b}\left( y\right)
f^{bdf}C^{d}\left( y\right) C^{f}\left( y\right) C^{+a}\left( x\right)
\right\rangle \,.  \label{A.6.6}
\end{eqnarray}

This identity has been verified by an explicit calculation in quantum
electrodynamics (within a gauge in which particles are free) as well as in
the first perturbative order of a non-Abelian theory.

It should be noted that the process of deriving the Ward identities with the
help of the usual procedure of a non-local change of variables in the
functional integral \cite{11} for the generating functional leads to a much
more involved expression than (\ref{A.6.6}). It is interesting, however,
that there exists another change of variables (\emph{supertransformation}),
that allows one to obtain (\ref{A.6.6}), as well as the usual Ward identity.

Consider the generating functional%
\begin{equation}
Z=\int dA\,d\varphi \,d\psi \,d\overline{\psi }\,dC\,dC^{+}e^{iL+i\widehat{Q}%
}\,,  \label{A.6.7}
\end{equation}%
where the expression for $L$ is given by formula (\ref{2.1}). Let us subject
(\ref{A.6.7}) to the change of variables (\ref{2.11})--(\ref{2.11'}), and
let us choose the gauge parameter as follows:%
\begin{equation}
\Lambda ^{a}\left( x\right) =\mu C^{a}\left( x\right) \,,  \label{A.6.8}
\end{equation}%
where $\mu $ is an anticommuting object. Notice that the choice (\ref{A.6.8}%
) leads to an exact form of transformations (\ref{2.11})--(\ref{2.11'}),
since $\mu ^{2}=0$. Besides, let us make a change of the field $C^{+}$,
namely,%
\begin{equation}
C^{+a}\rightarrow C^{+a}-\mu \alpha _{ab}t^{b}\left( x\right) \,.
\label{A.6.9}
\end{equation}%
The corresponding variation of the Lagrangian reads%
\begin{equation}
\delta L=\int dx\,C^{+a}\delta T^{ab}C^{b}=-\frac{\mu g}{2}\int
dx\,C^{+a}T^{ab}f^{bdf}C^{d}C^{f}\,,  \label{A.6.10}
\end{equation}%
where $\delta T^{ab}$ is the result of taking a variation of $T^{ab}$, whereas a
transition from the first equality to the second one in (\ref{A.6.10}) is
compensated by the following transformation of the field $C$:%
\begin{equation}
C^{a}\left( x\right) \rightarrow C^{a}\left( x\right) -\frac{g}{2}%
f^{adb}\Lambda ^{d}\left( x\right) C^{b}\left( x\right) \,,  \label{A.6.11}
\end{equation}%
where $\Lambda $ is given by formula (\ref{A.6.8}). As a result, we find
that Lagrangian (\ref{2.1}), or, more exactly, the action, is invariant
under (super)transformations (\ref{2.11})--(\ref{2.11'}), (\ref{A.6.11})
with the parameter $\Lambda $ defined by formula (\ref{A.6.8}), whereas the
Jacobian of this change equals to $1$. Therefore, variation affects only the
term with the sources, and so we obtain the following relation:%
\begin{equation}
\int dA\ldots \int dy\left[ \frac{g}{2}\theta
^{b}f^{bdf}C^{d}C^{f}+Q_{R}^{b}C^{b}-t^{b}\alpha _{bc}\theta ^{c}\right]
e^{iL+i\widehat{Q}}=0\,.  \label{A.6.12}
\end{equation}%
Differentiating (\ref{A.6.12}) by $\delta /\delta \theta ^{a}\left(
x\right) $, we obtain the Ward identity in the form (\ref{A.6.6}).

\end{document}